\documentclass[12pt]{JHEP3}

\usepackage{amsmath,epsfig}
\usepackage{amssymb,amsfonts}
\usepackage{latexsym}
\usepackage{slashed}
\usepackage{empheq}
\numberwithin{equation}{section}
\usepackage{dsfont}
\usepackage{cancel}
\usepackage{overpic}
\usepackage{subcaption}
\usepackage{subeqnarray}
\usepackage{xcolor}

\usepackage{graphicx}
\usepackage{longtable}
\usepackage{color}
\usepackage{bbm}

\newcommand{\exclude}[1]{}
\def\<{\langle}
\def\>{\rangle}

\def\a#1{\alpha_{#1}}

\def\beq{\begin{equation}}
\def\eeq{\end{equation}}
\def\be{\begin{equation}}
\def\ee{\end{equation}}
\def\bea{\begin{eqnarray}}
\def\eea{\end{eqnarray}}
\def\bal{\begin{align}}
\def\eal{\end{align}}

\def\2b2[#1,#2][#3,#4]{\left( \begin{array}{cc} #1 & #2 \\ #3 & #4 \end{array}
\right)}
\def\3b3[#1,#2,#3][#4,#5,#6][#7,#8,#9]{\left( \begin{array}{ccc} #1 & #2 #3 \\
#4 & #5 & #6\\#7&#8&#9\end{array} \right)}

\newcommand\fverb{\setbox\pippobox=\hbox\bgroup\verb}
\newcommand\fverbdo{\egroup\medskip\noindent%
                        \fbox{\unhbox\pippobox}\ }
\newcommand\fverbit{\egroup\item[\fbox{\unhbox\pippobox}]}

\newcommand{\bear}{\begin{eqnarray}}

\newcommand{\eear}{\end{eqnarray}}

\newcommand{\bsea}{\begin{subeqnarray}}
\newcommand{\esea}{\end{subeqnarray}}
\newbox\pippobox

\newcommand{\ud}{\mathrm{d}}

\def\6{\partial}
\def\a{\alpha}

\def\pa{\partial}

\def\e{\epsilon}
\def\m{\mu}
\def\n{\nu}
\def\r{\rho}
\def\s{\sigma}

\def\sp{\;\;\;,\;\;\;}

\def\sq
\def\a{\alpha}

\def\hri#1#2{\href{http://arxiv.org/abs/#1}{[ArXiv:#1]#2}}
\def\hre#1#2{\href{http://arxiv.org/abs/#1/#2}{[ArXiv:#1/#2]}}

\def\hrj#1#2{\href{www.doi.org/#1}{#2}}

\def\e{\epsilon}

\def\dd{{\rm d}}

\title{Critical scaling of the AC conductivity and momentum dissipation}
\author{N. Angelinos$^{\sharp}$, E.~Kiritsis$^{\flat,\natural}$, F. Pe\~na-Benitez$^\star$ \\
~\\
$^\natural$ \href{http://www.apc.univ-paris7.fr}{Universite de Paris, CNRS, Astroparticule et Cosmologie,  F-75006 Paris, France},\\
~\\
$^\flat$ \href{http://hep.physics.uoc.gr}{Crete Center for Theoretical Physics}, Institute for Theoretical and Computational Physics,
Department of Physics,  P.O. Box 2208,\\
University of Crete, 70013, Heraklion, Greece\\
~\\
$^\star$ Max-Planck-Institut  f\"ur Physik komplexer Systeme, N\"othnitzer Str. 38, 01187 Dresden, Germany\\
~\\
$^\sharp$ Department of Physics, University of Kentucky, Lexington KY, United States
}

\preprint{CCTP-2020-08\\
ITCP-IPP 2020/08}

\abstract{The scaling of the AC conductivity in quantum critical holographic theories at finite density, finite temperature and in the presence of momentum dissipation is considered. It is shown  that there is generically an intermediate window of frequencies in which the IR scaling of the AC conductivity is clearly visible.

}

\keywords{AC conductivity, Scaling, Strange metals, High-Tc superconductivity, holography}

\begin{document}
\maketitle

\section{Introduction and summary}

Critical behavior, scaling and universality are landmarks that stand out from the messy reality of materials.
It is unlikely in the space of possibilities that such behavior occurs at weak coupling, but it is possible and examples exist.
Typically, however, such behavior happens at moderate or strong coupling, and then inferring the physics becomes difficult.

A concrete example where quantum criticality, and scaling, may be realized, is in strange metals, that include high-T$_c$ superconductors, \cite{ander,zaanen2,zaanen3,zaanen4,Hussey}.
The scaling of the DC conductivity in the cuprates, has been since the start, one of the basic hallmarks, exhibiting linear resistivity at optimal doping over a large range of temperatures, \cite{Ong}-\cite{Legros}. More recent and refined data on the DC conductivity in clean materials, and a novel parametrization indicated an unusual low-temperature asymptotic behavior and the existence of a line of quantum critical points, \cite{anomalous}.  Similar evidence for a line of critical points was found in other cases, \cite{line1,line2}.

The magnetoresistance  of related materials has indicated also exotic behavior and a different scaling of the Hall angle in the overdoped regime, \cite{Kendziora1992}-\cite{Ando}.  New measurements of the magnetoresistance in strong magnetic fields, have also indicated scaling both in pnictides, \cite{magneto1,magneto11}, and cuprates, \cite{magneto2}.

Studies of the AC conductivity, \cite{scal1}, have also produced scaling in $\omega$, in an intermediate range of frequencies. Although, it is typical for critical theories to induce a scaling in the AC conductivity in the far IR, it is unusual that such scale should survive and be visible at higher frequencies.
Further experiments have shown similar scaling of the AC conductivity in other strange metals, \cite{scal2,scal3}.

Despite the wealth of experimental indications of scaling phenomena, progress in theory has been slower. The main reason is that few scale invariant quantum critical points were known in two spatial dimensions and none that is non gaussian. This has changed with the AdS/CFT correspondence, \cite{Malda}, also known as holography. This emergent in string theory, provided a   valuable tool in analysing quantum field theories at strong coupling and was especially successful in describing  scale invariant, strongly coupled theories at finite density. This has led to a classification of quantum critical behavior at string coupling, as a function of the symmetries, \cite{cgkkm}-\cite{helical}.
Moreover, it provides techniques for calculating the quantum effective potential at finite density and temperature, \cite{Veff}, that provides a powerful tool in studying phase transitions.
A wide spectrum of condensed matter problems was addressed using these techniques, and this progress is summarized in reviews and lectures, \cite{Hartnoll}-\cite{HLS}, as well as in books, \cite{Zbook}-\cite{book}.
A recent overview of the progress in the field can be found in \cite{Zaanen}.

In the context of holography, studies have indicated that the behavior of fermions and their correlations at strong coupling may be radically different from that at weak coupling, \cite{liu,CZS}. This gave in particular, a class of realizations of the marginal Fermi liquid, \cite{liu}, realizing correlators that were associated with the linear behavior of the DC conductivity, \cite{marginal}.
Further holographic studies have analysed the constraints on the realization of a linear in $T$ DC conductivity and provided explicit examples, \cite{HPST,KKP,Pal}. In particular, in \cite{KKP} the model exhibits non-relativistic $z=2$ Schrodinger symmetry, and reproduces the $T+T^2$ behavior of the DC conductivity, \cite{anomalous}, and a Hall angle and magnetoresistance that are in agreement with data at very low temperatures, \cite{mackenzie}. Moreover, it realized the idea of a line of quantum critical points, suggested in \cite{anomalous} and predicted scaling relations in th presence of the magnetic field.

The scaling properties in the presence of magnetic fields are nicely described by holographic critical theories. In general charge dynamics is described by the DBI action, and its scaling analysis in the presence of magnetic fields has yielded a rich collection of scaling behaviors, \cite{KKP,DBI}.
In one particular case, such a scaling behavior matches the one seen in the pnictides and cuprates, \cite{magneto1,magneto11,magneto2}.

The scaling of the AC conductivity observed in \cite{scal1} has been tougher to crack. A conventional approach suggested that it may arise from the interaction of the fermions with a Bose sector other than the phonons, \cite{scalch}.
 In holography it has been studied since
\cite{cgkkm} observed that in scaling critical geometries, there is also a scaling AC conductivity and computed the scaling exponent, as a function of the other critical exponents, for the case where the EM gauge boson is the same as the one that seeds the scaling IR geometry. This issue  was further studied in \cite{G}.
It was then shown that one could also have intermediate regimes which have also scaling properties, and they may affect the intermediate scaling of the AC conductivity, \cite{GC}.

A definitive study has been done in \cite{scaling}. There, the model of \cite{KKP} was analyzed by computing its AC conductivity in its various regimes. A related study of scaling was done in \cite{KP}.

 \subsection{On the scaling of the holographic AC conductivity}

The first theory that was analyzed in \cite{scaling} was the holographic DBI theory of a strange metal, proposed in \cite{KKP}.
This theory has several parameters, but the physics depends only on two scaling variables, $t$ that is proportional to the temperature, and $J$ that is proportional to the charge density. They both take all positive real values. The doping parameters is part of both scaling variables.

The $T+T^2$ behavior of the resistivity
in \cite{anomalous} and the $T+T^2$ behavior of the inverse Hall angle, observed in \cite{mackenzie} at {\em very low temperatures} $T<30K$, where a single scattering rate is present, were successfully described in this theory.
The model   is also in accord with the distinct origin of the criticality at very low temperatures advertised in \cite{Hussey2},
while the higher temperature, $T>100K$, scaling has different behaviors between
the linear temperature resistivity and the quadratic temperature inverse Hall angle, signaling two
scattering rates \cite{Tyler1997}. This regime in the model is different however from what experiments show about the cuprates.

In addition to the resistivity and inverse Hall angle, very good agreement was  also found with experimental
results of the Hall Coefficient, magnetoresistance and K\"ohler rule on various high-$T_c$ cuprates,
\cite{anomalous},\cite{Kendziora1992}-\cite{Ando}.
The model provided also a  change of paradigm from the notion of a quantum critical point, as it is quantum critical at $T\to 0$ on the entire overdoped region as suggested by the data of \cite{anomalous}.
The DC conductivity of this theory, as is usual in quantum critical theories, has two contributions, \cite{KKP}. One, that we call the Drude contribution,  is related to momentum dissipation in the standard fashion, although here there are no weakly coupled quasiparticles. The other is independent of the charge density and is the Quantum Critical contribution.

There are two main regimes on the $(t,J)$ plane.
 They are best described by a parameter $q$ that is a function of $t,J$ and  distinguishes between the two regimes. When $q\gtrsim 1$ the DC conductivity is dominated by the Drude (drag) contribution. When  $q\lesssim 1$ the DC conductivity is dominated by the Quantum Critical contribution.
As the drag contribution to the conductivity is proportional to charge density, it follows that at zero charge density ($J=0$) we are always in the QC/PP regime. These two regimes are shown in figure \ref{fig_T00}.

\begin{itemize}

\item In the Drude regime ($q\gg 1$), when $t\ll 1$ the resistivity is linear in $t$ (and consequently in the temperature). This is the {\em linear regime}. When $t\gg 1$, the resistivity is quadratic in $t$.  This is  the {\em quadratic regime}.

\item In the Quantum Critical   regime,  ($q\ll 1$), when $t\ll 1$  the resistivity behaves as $\rho\sim t^{-{3\over 2}}$. This is {\it regime I}.
    When $t\gg 1$ the resistivity behaves as $\rho\sim t^{-{1\over 2}}$. This is  {\em regime II}.

\end{itemize}

In the $t\to 0$ limit the theory has an effective Lifshitz exponent $z=2$ while as $t\to\infty$ it crosses over to an effective relativistic Lifshitz exponent, $z=1$, \cite{KKP}.
What we find in our analysis is as follows:

\begin{enumerate}

\item A {\em generalized relaxation time} $\tau$ can be defined by the IR expansion of the AC conductivity,
     \be
     \sigma(\omega) \approx \sigma_{DC}\left(1+ i ~\tau~\omega + \mathcal{O}(\omega^2)\right)
\label{p1i}\ee
In the presence of a Drude peak, this is the conventional definition of an associated relaxation time. When there is no Drude peak present, $\tau$ is still well-defined, although in that case the interpretation as a relaxation time is lost.

   In \cite{scaling}   an analytical formula for $\tau$ was given.
    It simplifies for large and small values of the scaling temperature variable $t$.  In the regime I
\be
\tau\sim \sqrt{t}
\label{p2i}\ee
while in the regime II (with $t\gg 1$) $\tau$ is set by the inverse of the temperature
\be
\tau\simeq \frac{1}{ t}
\label{p3}\ee

\item In the Drude regime ($q\gg 1$) where the dominant mechanism for the conductivity is momentum dissipation, there is a clear Drude peak in the AC conductivity.

    In the Quantum Critical  regime there is no Drude peak and we have an incoherent  AC conductivity.

\item At zero charge density (this is the Quantum Critical regime) there is a scaling tail for the AC conductivity that behaves as
    \be
    |\sigma|\sim \left({\omega\over t_{eff}}\right)^{-{1\over 3}}\sp Arg(\sigma)\simeq {\pi\over 6}
\label{p4}\ee
For finite charge density, this tail survives not only in the Quantum Critical  regime  but also in parts of the Drude regime.

\item This scaling tail of the AC conductivity generalizes to more general scaling holographic geometries, as previously described in \cite{cgkkm,G}. In this case, the theory was taken to have $T=0$ and no momentum dissipation.

  In particular, for a metric with Lifshitz exponent $z$, hyperscaling violation exponent $\theta$ and conduction exponent $\zeta$, \cite{GK2,G} with $d$ spatial boundary dimensions,  we find that in general
   \be
    |\sigma|\sim {\omega}^{m}\sp Arg(\sigma)\simeq -{\pi~m\over 2}
\label{p5}\ee
with
\be
 m = \left|{z+\zeta-2\over z}\right|-1\,,
\label{p6}\ee
There are several constraints in the parameters of this formula that are detailed in \cite{scaling}.
This formula is valid when the associated charge density does not support the IR geometry. In this case the scaling exponent can become negative but is also $m\geq -1$.

\begin{figure}[t]
	\begin{center}
		\includegraphics[scale=0.49]{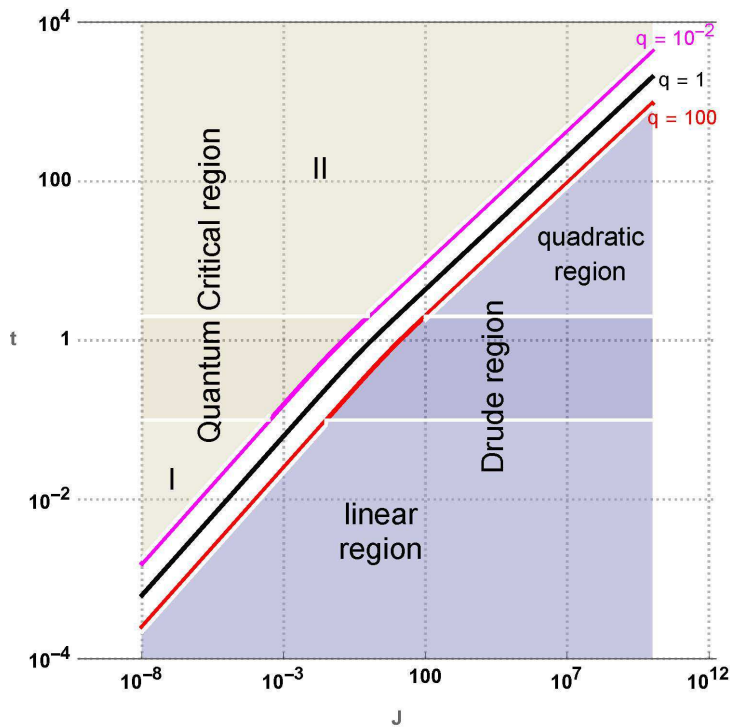}
	\end{center}
	\caption{The parameter landscape of the DBI Theory and the regimes for the DC conductivity}
	\label{fig_T00}
\end{figure}

\item In the special case where the associated gauge field seeds the IR scaling geometry,  the exponent $m$ takes the value
    \be
    m=\Big|{3z-2+d-\theta\over z}\Big|-1
   \label{p7} \ee
and is always positive.

\item An important question  is whether the scaling of the AC conductivity described above, for the general scaling geometries, is controlled by the dynamics of the charge density, or it is decided by the neutral system.
    What was found in \cite{scaling} is that it is the neutral system that decides the exponent $m$. The charges contribution is almost always subdominant.

\end{enumerate}

The results of \cite{scaling} have positively indicated that holographic Quantum critical theories at finite density and $T=0$, in the absence of momentum dissipation have a scaling IR AC conductivity, roughly of the type seen in experiments.

There is however, an important difference, with the AC conductivity seen in experiments as one needs to turn on non-zero temperature and momentum dissipation and see to what extend this scaling survives these effects. It must also appear in the mid-frequency range roughly $T\ll \omega\ll \mu$.

\subsection{Results and Outlook}

In this paper we take the first step towards understanding how the scaling of the AC conductivity  survives the effects of temperature and momentum dissipation.
 The example we analyze is the simplest possible one: the UV theory is a Conformal Field Theory\footnote{The UV theory is assumed to be Lorentz invariant. The charge density breaks the Lorentz-invariance. The IR theory has an emergent scale invariance in time only.}  (CFT). The holographic theory this is associated with the four-dimensional Anti-de Sitter geometry (AdS$_4$). The IR theory is also a CFT but dominated by charge density and has scale invariance in time only. It is associated with the two-dimensional Anti-de Sitter geometry,  (AdS$_2$).

 There is a characteristic scale that controls the passage from AdS$_4$ to AdS$_2$ (at T=0) and this is the charge density (or chemical potential $\mu$).
We also add momentum dissipation. This introduces a new characteristic (mass) scale, $k$ that competes with the charge density.
We have in total three scales, $\mu$, $T$ and $k$.
Therefore this theory depends on two dimensionless ratios, $\tau$ and $\kappa$ that control the importance of temperature and momentum dissipation,
\be
\tau=2\pi{T\over \mu}\sp \kappa \equiv {k\over \mu}
\label{p8}\ee
Moreover, we define the following quantity, related to momentum dissipation
\be \lambda\equiv\sqrt{k\over 2\mu}=\sqrt{\kappa\over 2}.\label{lambda}\ee

The theory at $\tau=\kappa=0$ has two scaling regimes. In the UV  regime the scaling exponent is $m=0$ ($z=1$, $\zeta=0$ in (\ref{p6}). It is a well known result that in a Lorentz-invariant and scale invariant theory at zero density, in $d=2$ the conductivity is a (dimensionless) constant. We normalize, without loss of generality this constant to one.
In the IR,  $m=2$ ($z=1$, $d=2$, $\theta=2$ in (\ref{p7}).

We generically find four distinct regimes.
\begin{itemize}

\item The Drude regime for
$${\omega\over \mu}\ll \lambda\;.$$
 In this regime momentum dissipation produces a Drude peak\footnote{As before, this is despite the fact that there are no quasi-particles in the holographic theory.}.

\item The temperature dominated regime:
$$\lambda \ll {\omega\over \mu}\ll \tau\;.$$
 Here thermal effects dominate the AC conductivity.

\item The scaling (intermediate) regime:
$$\max\left\{\tau,\lambda\right\}\ll {\omega\over \mu}\ll 1\;.$$
In this regime, if it exists, the AC conductivity is showing its $\omega^2$ IR scaling, unmasked. It exists only if $\max\left\{ \tau,\lambda \right\}$ is sufficiently smaller that one.

\item The UV regime,
$$
{\omega\over \mu}\gg 1\;.
$$
In this regime the AC conductivity is that of the UV theory, ie, a constant. It has been normalized to 1.

\end{itemize}

The four regimes along with the real part of the AC conductivity are summarized in the table below.

We conclude that our expectations are verified in the simple example that has been analyzed here. The next step is a choice of theory with non-trivial scaling exponents that provide a negative value for the AC exponent $m$, and study similarly the effects of temperature and momentum dissipation on the visibility of scaling of the AC conductivity.
Moreover, this mechanism must be implemented in the more complex context of a theory which contains the competition of phases, giving rise to the strange metal phase diagram, along the lines of \cite{competition}.

\begin{center}
	\begin{tabular}{|c|c|c|}
		\hline
		Regime & frequency & $Re[\sigma]$ 
		\\
		\hline\hline
		{Drude} & $ {\omega\over \mu}\ll\lambda$ & ${\kappa^2 \over \kappa^4+12\omega^2/\mu^2}$ 
		\\
		\hline
		{temperature-dominated} & $\lambda \ll {\omega\over \mu}\ll \tau$ & ${1\over 3}{ \tau}^2$ 
		\\
		\hline
		{scaling} & $\max\left\{ \tau,\lambda \right\}\ll {\omega\over \mu}\ll 1$ & ${1\over 3}\left(\frac{\omega}{\mu}\right)^2$ 
		\\
		\hline
		{UV} & ${\omega\over \mu}\gg1$& $1$ 
		\\
		\hline
	\end{tabular}
\label{tab}\end{center}

The paper is organised as follows, In Sect. \ref{Sect_RNAdS2} we introduce the holographic model in consideration and discuss its black-hole solution and equilibrium properties. In Sect. \ref{Sect_AC} we compute the frequency dependence of the conductivity in three different regimes, the first case corresponds with the momentum conserving zero temperature case. Then, we switch temperature on and study the conditions for scaling conductivities in the IR regime. As a last step we include relaxation of momentum and study all the possibilities in the IR conductivity. We close our analysis with section \ref{concl}, where our conclusions are presented  and where we discuss possible generalizations.

\section{The Reissner-N\"ordstrom black-hole and the AdS$_2$ IR-scaling asymptotics}\label{Sect_RNAdS2}

We consider a 2+1 (scale invariant) Conformal Field Theory (CFT) with a (global) conserved U(1) charge in a flat Minkowski space-time. The conserved charge allows us to consider the theory at finite charge density, a context relevant for addressing many-body problems.
Our scale invariant Quantum Field Theory (QFT) is not a generic relativistic theory: it is a large-N theory at nearly infinite coupling constant. Here $N$ is the number of colors and unlike other Large-N examples used in condensed matter physics, it includes an $SU(N)$ gauge interaction that makes the theory much more complex.

Such large-N gauge theories at strong coupling are known as ``holographic", as they are dual to gravitational theories in higher dimensions (typically one-higher dimension) on nontrivial geometric spaces that are asymptotic to Anti De Sitter space. In our example, the dual gravitational theory will have 3+1 dimensions. The (3+1)-dimensional spaces that are relevant, have always a boundary that has the same geometry as the space on which the QFT lives. In our case the boundary will be flat (2+1)-dimensional Minkowski space.
The AdS/CFT correspondence and the associated applications to condensed matter problems are treated in several extensive references, \cite{Hartnoll}-\cite{book}.

As a starting point, we introduce the gravitational action for the system in consideration, which consists of a (bulk)\footnote{In this work, by bulk we mean the (3+1)-dimensional space where the gravitational theory lives, whereas the boundary is (2+1)-dimensional and is the space on which the dual QFT lives.}  gravitational  Einstein-Maxwell theory in $3+1$ space-time dimensions, coupled to a set of axion fields\footnote{In this work, we refer to axion fields, as scalar bulk fields without a potential.}  responsible for the non-conservation of momentum in the boundary field theory,
 \cite{mr1}-\cite{insu}.

Einstein-Maxwell theory in 3+1 dimensions is the holographic (bulk) description of the universal sector of a holographic (2+1)-dimensional CFT. It contains a metric, $g_{\m\n}$,  dual to the energy-momentum tensor of the CFT, and a gauge field, $A_{\m}$,  dual to the conserved U(1) current.
We have also introduced a source of momentum dissipation in the theory, that is generated by two scalar fields, $\phi_{1,2}$, without a potential. Such fields, that we call axions, implement a source of momentum dissipation in the continuum limit.

The action for the bulk theory is
\be \label{eq:bulkaction}
S= \frac{1}{16 \pi G_N} \int \dd ^4 x \sqrt{-g} \left( R+\frac{6}{L^2}-\frac{L^2}{4}F_{\m\n}F^{\m\n}  -\frac{1}{2}\sum_{n=1}^2 ~\partial_{\m} \phi_n\pa^{\m}\phi_n    \right)\,.
\ee
where $R$ is the Ricci scalar of the bulk metric and
\be
F_{\mu\n}=\pa_{\m}A_{\n}-\pa_{\n}A_{\m}
\label{1}\ee
is the field strength of the bulk gauge field.

The equations of motions are derived in appendix \ref{app:EoM}.
Solutions to the equations of motion which are asymptotically AdS, and have specific boundary conditions at the AdS boundary, are interpreted as saddle points of the CFT.

 The theory  in (\ref{eq:bulkaction}) admits a charged black-hole solution of the form
\be
\dd s^2= \frac{L^2}{r^2}\left(-f(r) \dd t^2 +\dd x^2+\dd y^2+\frac{1}{f(r)}  \dd r^2\right)\,,
\label{2a}\ee
\be
A_t=\psi(r)\sp
\phi_1 = k\, x\sp \phi_2 = k\, y\,,
\label{2}\ee
where
\be
 f(r)=1+{1\over 4}q^2 r^4-{1\over 2} k^2 r^2-m{r^3}\,,
 \label{3a}\ee
 \be
 \psi(r)=\mu-q\, r\,,
\label{3}\ee
and $q,m$ are proportional to the charge and energy density of the system respectively.\footnote{This black-hole solution is the saddle point that described the ground state of the theory at finite temperature and U(1) charge.}
Here, $x,y$ are cartesian coordinates in space, $t$ is the time, and $r$ is the holographic coordinate. The boundary of the bulk space is at $r=0$.

The two scalars have linear solutions that break translational invariance in the spatial directions and therefore provide a source of momentum dissipation.
We have chosen the parameters of that solution so that rotational invariance remains intact (mostly for simplicity). We can generalize this to a solution where momentum dissipation is different along different spatial directions. The  regularity condition at the horizon implies that
\be
q={\mu\over r_0}\;,
\label{4} \ee
 where $r_0$ is the horizon radius.

 At this point, it is important to review the parameters that enter in the solution\footnote{We use units where $c=\hbar=1$.}. First of all the gravitational action has two dimensionfull parameters. One is the bulk Newton's constant $G_N$ and the other is the AdS curvature length $L$. The dimensionless number is
\be
{L^2\over G_N}\sim N^2~~\gg ~~1
\ee
where $N^2$ is the number of adjoint degrees of freedom of the quantum field theory. Only $L$ enters the solution.
The rest of the parameters involve:

$\bullet$ The mass of the black-hole, $m$,  which gives the energy of the canonical ensemble.

$\bullet$ The charge $q$,  with dimension of mass$^2$, that determines the charge density of the dual CFT.

$\bullet$ The  parameter $\mu$ with dimension of mass, that determines the chemical potentials of the dual CFT. It is related to the charge density via the relation in (\ref{4}).

$\bullet$ The parameter $k$, with dimension of mass (or inverse length), that controls the breaking of translation invariance and therefore the rate of momentum dissipation in the system.

$\bullet$ The temperature $T$ of the ensemble is fixed and related to the other parameters in a way we shall describe below.

 The horizon radius $r_0$ is related also to other parameters of the solution. In order to fix the value of $r_0$ in terms of the physical scaling parameters $T,\mu,k$ we need to solve for the condition $f(r_0)=0$, where $r_0$ is given by the smallest positive solution\footnote{Notice that the  boundary is sitting at $r=0$, and the outer horizon corresponds with the smallest positive solution of $f(r_0)$.} of the polynomial in $x$
 \begin{equation}\label{eq_polyhor}
1+{1\over 4}q^2 x^4-{1\over 2} k^2 x^2-m x^3=0\,,
\end{equation}
For $q\ne 0$, the polynomial \eqref{eq_polyhor}, always has exactly two positive real roots $r_1,r_2$. As long as the following inequality is satisfied
\be
108m^2>k^2\left(k^4-36q^2\right)+\left(k^4+12q^2\right)^{3/2} ,\label{bhmc2}
\ee
the other two roots are complex. If (\ref{bhmc2}) is not satisfied, the other two roots are also real, but negative, hence they do not affect us, since $r$ is nonnegative in our coordinate system.
We can now factorize the blackening factor as follows
\be
f(r)=\left(1-{r\over r_1}\right)\left(1-{r\over r_2}\right)\left(1+{r_1+r_2\over r_1r_2}r+{r_1^2+r_1r_2+r_2^2\over r_1^2r_2^2}r^2\right)\,,\label{eq_factf}
\ee
where the two positive real roots $r_1,r_2$ satisfy
\begin{equation}\label{eq_qkmvsr1r2}
{1\over 2}k^2 +{1\over 4}{q^2}r_1r_2={1\over r_2^2}+{1\over r_1r_2}+{1\over r_2^2}\,,\qquad m={r_1^3+r_1^2r_2+r_1r_2^2+r_2^3\over r_1^3r_2^3}.
\end{equation}
Finally we identify $r_0=\min (r_1,r_2)$ with the black-hole horizon, and $r_\star=\max (r_1,r_2)$  as the interior (Cauchy) horizon\footnote{Notice that in our coordinates the boundary sits at $r=0$.} characteristic of charged black-holes.
The Hawking temperature for such a black-hole reads
\be
T={1\over 4\pi r_0}\left(3-{1\over 2}k^2r_0^2-{1\over 4}{q^2r_0^4}\right)\label{eq_temp},
\ee
which can be shown  to be positive definite, after using the left equation of Eq. \eqref{eq_qkmvsr1r2} to construct the following relation
\be
{1\over 2}k^2r_0^2+{q^2r_0^4\over 4}\leq{1\over 2}k^2r_0^2+{q^2r_0^3r_\star\over 4}={r_0^2\over r_\star^2}+{r_0\over r_\star}+1\leq 3.\label{ineq1}
\ee
The inequality is saturated at extremality ($r_0=r_\star$) where the temperature vanishes. In addition, the black-hole mass can be written as
\be
m=\frac{4-2r_0^2k^2+r_0^2\mu^2}{4r_0^3}\label{bhm1}.
\ee
Considering we are interested in exploring the low temperature properties of the geometry and the optical conductivity, it is convenient to introduce the scaling (dimensionless) variables
\be
\tau = 2\pi {T\over \mu}\sp \kappa ={k\over \mu}\,.
\label{6}\ee
In terms of the dimensionless temperature $\tau$ and momentum relaxing parameter $\kappa$, the black-hole horizon radius $r_0$  can be written as
\be
\mu r_0\equiv F(\tau,\kappa)=\frac{6}{\sqrt{6 \kappa ^2+4\tau ^2+3}+2\tau }\,.
\label{7}\ee
Therefore, the corresponding thermodynamic quantities, energy density $\e$, entropy density $s$ and charge density $q$, take the following form
\be
\mu^{-3}\e=2F(\tau,\kappa)^{-3}+{1-2\kappa^2\over 2}F(\tau,\kappa)^{-1}\sp  \mu^{-2}s=4\pi F(\tau,\kappa)^{-2} \sp  \mu^{-2}q= F(\tau,\kappa)^{-1}
\label{8}\ee

\subsection{The IR (near-extremal) AdS$_2$ geometry}

As it is well known, \cite{liu},  the zero temperature near-horizon geometry of this black-hole is $AdS_2\times \mathbb R^2$.
 To make this manifest, we  rewrite the blackening factor $f$ as follows
\be
f(R) = 2\tau F^2 \frac{R}{L^2\mu}+\frac{F^3  \left( (\kappa ^2+1)F-4 \tau \right)}{2 } \frac{R ^2}{\mu ^2 L^4}+\frac{F^4 \left(2 \tau -( \kappa ^2+2)F\right)}{3}\frac{R ^3}{ \mu ^3 L^6}+\frac{F^6}{4} \frac{R ^4}{ \mu ^4 L^8},
\ee
with
\be
R=\frac{L^2}{r_0^2}(r_0-r)
\label{zz} \ee
 and $F$ the function defined in (\ref{7}).  In terms of the new radial coordinate $R$, the metric reads
\begin{equation}
ds^2=\frac{L^6}{r_0^2(L^2-r_0 R)^2}\left(-f(R)~dt^2+\frac{r_0^4}{L^4}\frac{dR^2}{f(R)}\right)+\frac{L^6}{r_0^2(L^2-r_0R)^2}d\vec x^2\,.
\end{equation}
At zero temperature ($\tau=0$) and in the region where $R\ll\mu L^2$, the space-time is approximately AdS$_2\times \mathbb R^2$:
\begin{equation}\label{eq_AdS2}
ds^2=-\frac{R^2}{L_{IR}^2}dt^2+\frac{L_{IR}^2}{R^2}dR^2+\frac{L^2}{r_0^2}d\vec x^2\,,
\end{equation}
with the AdS$_2$ radius given by
\begin{equation}
L_{IR}^2=\frac{1+2\kappa^2}{1+\kappa^2}\frac{L^2}{6}\,.
\end{equation}
If $\tau$ is finite but small, in the region near the horizon
\be {R\over \mu L^2}\ll 1,
\ee
we obtain an $AdS_2\times \mathbb R_2$ black hole
\be ds^2=-g(R)dt^2+{dR^2\over g(R)}+{L^2\over r_0^2}d\vec x^2,\ee
where
\be g(R)={R^2\over L_{IR}^2}\left(1-{6+8\kappa^2\over (1+\kappa^2)\sqrt{3+6\kappa^2}}\tau-R^{-1}{2\tau\sqrt{3+6\kappa^2}\over 3(1+\kappa^2)}\right)+\mathcal O(\tau^2).\ee
In the intermediate region
\be
\tau {L_{IR}^2\over L^2}\ll {R\over \mu L^2}\ll 1,\label{ads2_reg}
\ee
we have
\be g(R)\approx \frac{R ^2}{ L_{IR}^2}\left(1-\tau{2(4\kappa^2+3)\over (1+\kappa^2)\sqrt{3+6\kappa^2}}\right)={R^2\over L_{IR}'^2},\ee
where we defined
\be L_{IR}'^2=L_{IR}^2\left(1+\tau {2(4\kappa^2+3)\over (1+\kappa^2)\sqrt{3+6\kappa^2}}+\mathcal O(\tau^2)\right).\ee
Therefore in the region (\ref{ads2_reg}) we still have an $AdS_2\times \mathbb R^2$ geometry
\be ds^2=-{R^2\over L_{IR}'^2}dt^2+{L_{IR}'^2\over R^2}dR^2+{L^2\over r_0^2}d\vec x^2.\ee
with a modification to the $AdS_2$ radius stemming from the blackening factor of the $AdS_2$ black hole.

\section{The AC conductivity and its critical IR scaling}\label{Sect_AC}

We now study the AC conductivity of our theory, and in particular its scaling form in appropriate frequency ranges.

In order to understand the conditions under which scaling tails appear in the system, we  consider first the zero temperature ($\tau=0$) and momentum conserving case ($\kappa=0$).

To compute the AC conductivity,  we introduce perturbations propagating on the extremal ($\tau=0$) Reissner-N\"ordstrom black-hole solution, and use linear response,  following the prescription introduced in \cite{Son:2002sd}.

 After studying the zero temperature case,  we turn-on  a small temperature in the system. In the  last step, in addition to the temperature,  we also include momentum relaxation.

\subsection{IR scaling of the AC conductivity}

\begin{figure}[t]
	\begin{center}
		\includegraphics[scale=0.49]{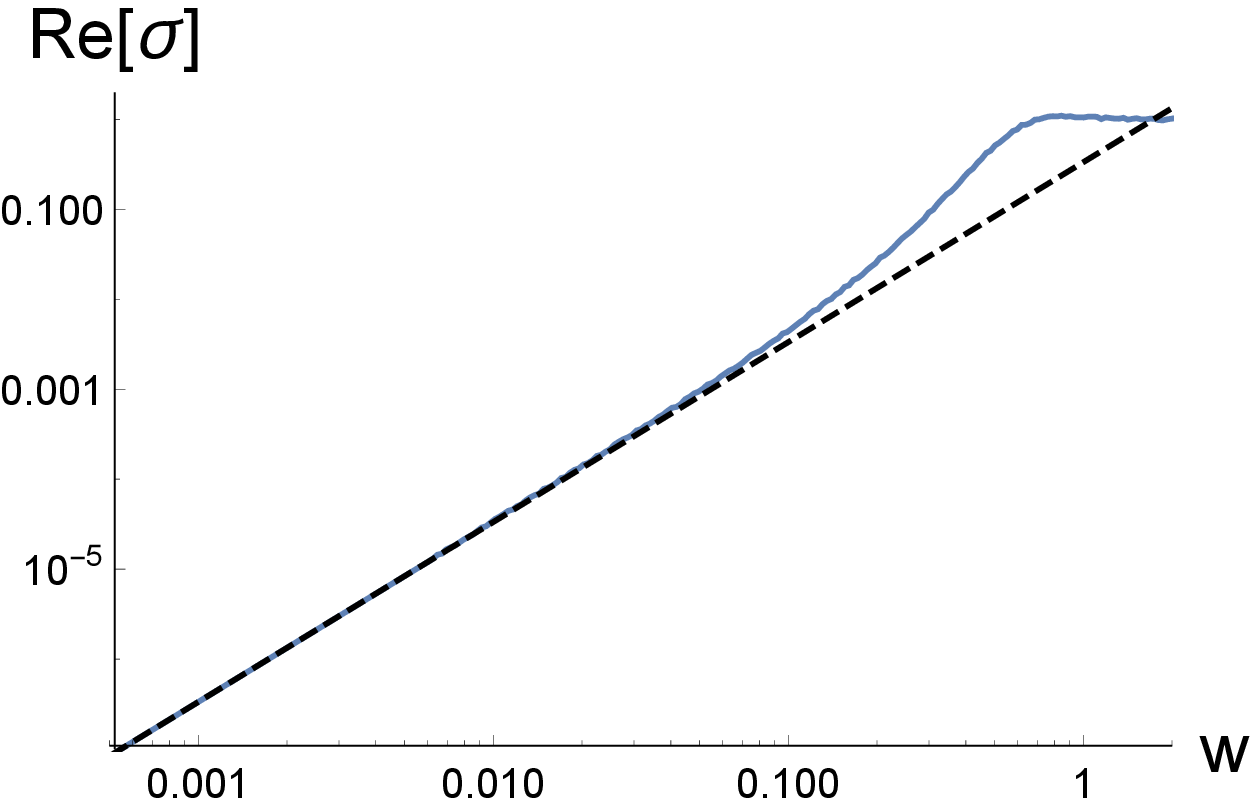}
		\includegraphics[scale=0.49]{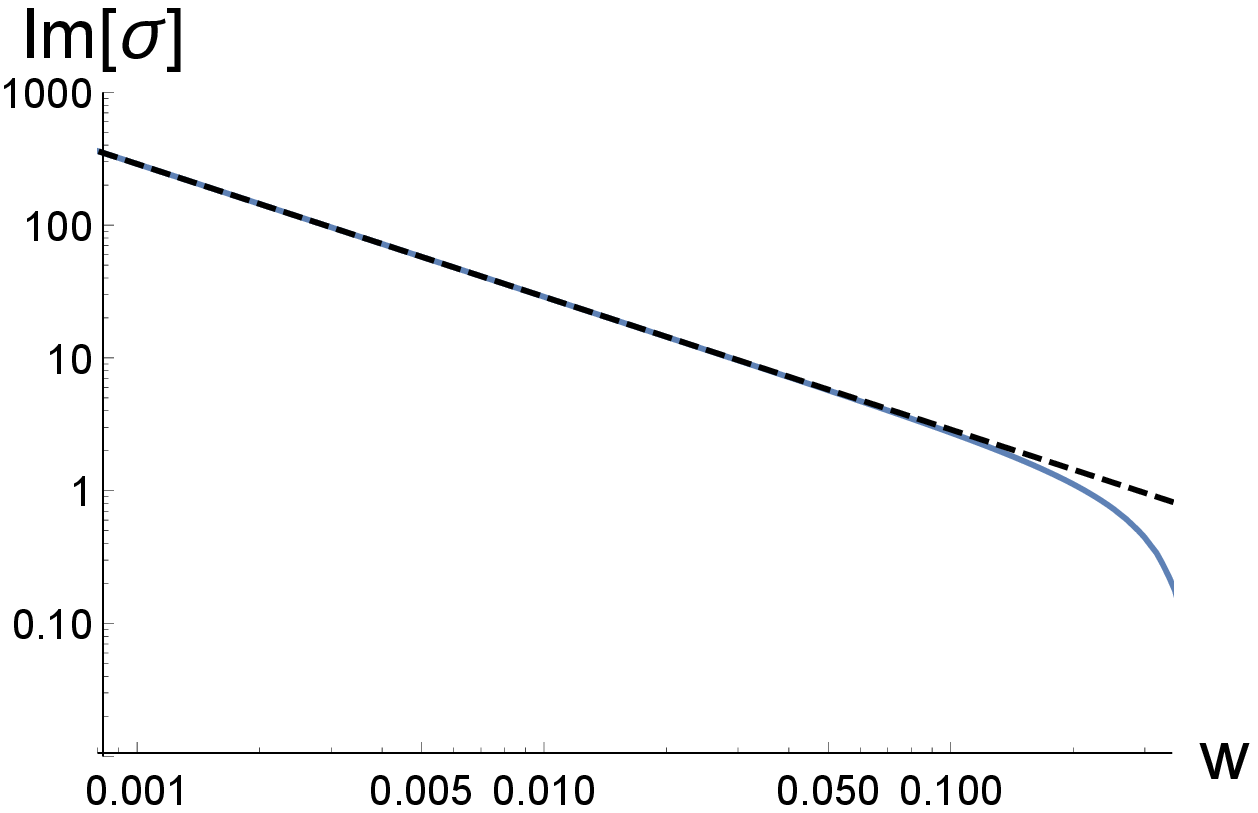}
	\end{center}
	\caption{Frequency dependence of the conductivity at $\tau=\kappa=0$. The real part (left) shows the scaling behaviour $\mathrm{Re}[\sigma]\sim w^2$. On the contrary, the imaginary part (Right) shows a  $\mathrm{Im}[\sigma]\sim w^{-1}$ divergence, indicating the presence of a $\delta(w)$ in the real part, as required by momentum conservation. The black dashed line corresponds to the fitting $\mathrm{Re}[\sigma]= w^2/3$ and $\mathrm{Im}[\sigma]= 1/(2\sqrt{3}w)$.}
	\label{fig_T0}
\end{figure}
\begin{figure}[t]
	\begin{center}
		\includegraphics[scale=0.49]{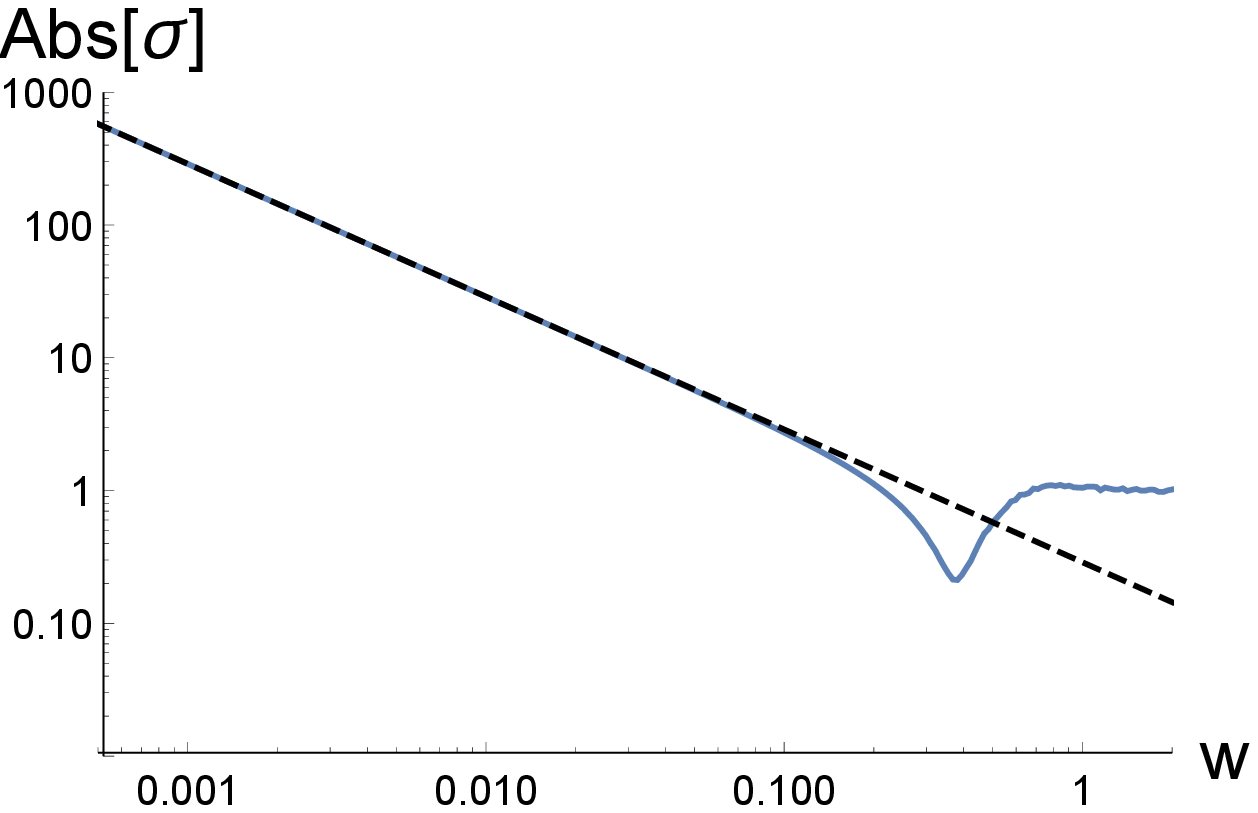}
		\includegraphics[scale=0.49]{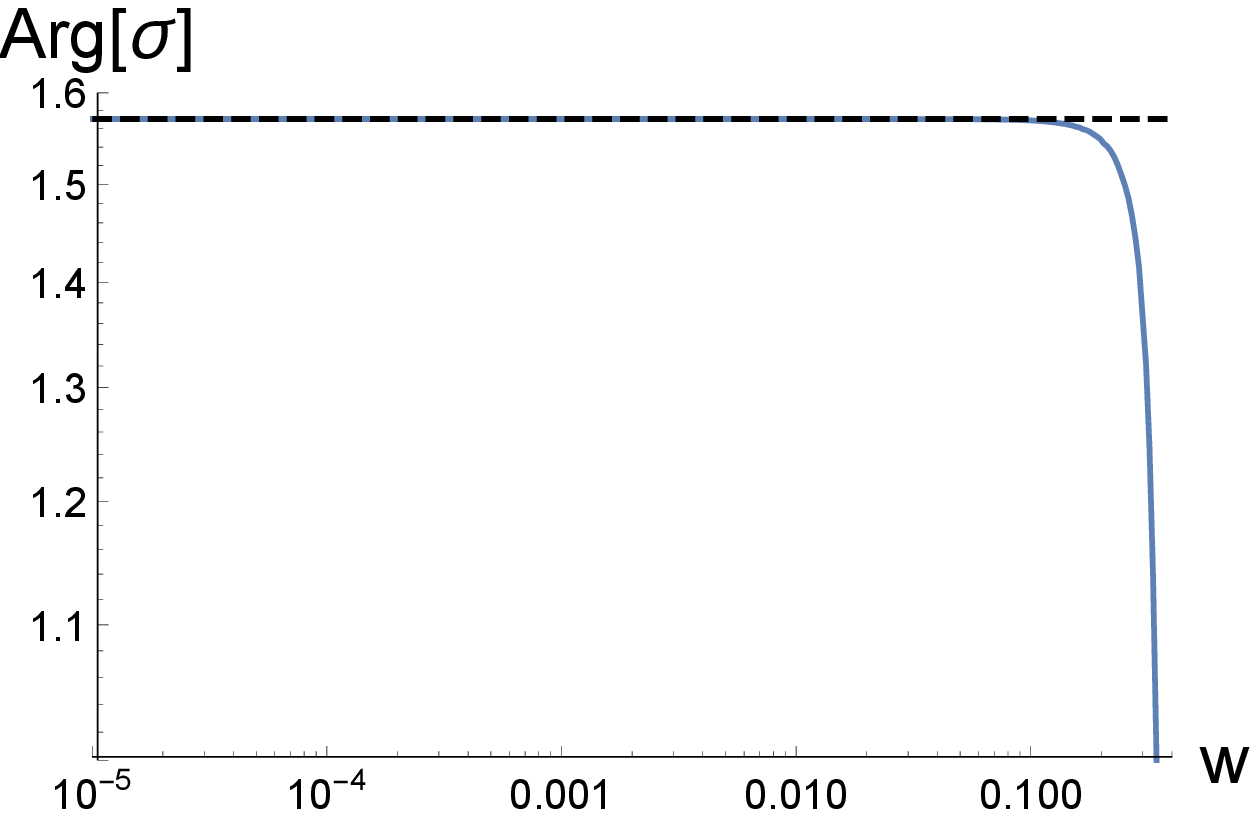}
	\end{center}
	\caption{Absolute value (Left) and argument of the conductivity (Right) at zero temperature momentum breaking parameter ($\tau=\kappa=0$). The dashed line shows a fitting with $|\sigma|={1\over 2\sqrt 3 w}$ and  $\arg[\sigma]={\pi\over 2}$ respectively. }
	\label{fig_T0_b}
\end{figure}

	
We begin with vanishing temperature and absence of momentum relaxation. The background is equivalent to the extremal AdS-Reissner-N\"ordstrom  black-hole  which has no independent dimensionless parameters. The fluctuation equation relevant for the computation of the electrical conductivity reads (see Appendix \ref{app_fluct} for the derivation)
\be\label{eq_flucteomT0}
fa_x'' + f'a_x' + 12\left(\frac{w ^2}{ f} -  \rho^2\right)a_x=0\,,
\ee
where the frequency is measured in units of chemical potential
\be
 w={\omega\over \mu}
\label{w} \ee
and
\be f(\rho)=(1-\rho)^2(1+2\rho+3\rho^2)\sp \rho=r/r_0\;.\ee
For the extremal  black-hole, the horizon is located at $(\mu r_0)^2=12$.  In particular, equation \eqref{eq_flucteomT0} has an irregular singular point at $\rho=1$, implying a near-horizon behaviour given by
\begin{equation}\label{eq_nearhorCond1}
a_x\sim (1-\rho)^{-4\sqrt 3/9iw } e^{\frac{i w }{\sqrt 3(1-\rho)}}.
\end{equation}
After numerically solving the differential equation with the near-horizon condition \eqref{eq_nearhorCond1}, the frequency dependence of the conductivity can be computed. The results are shown in Fig. \ref{fig_T0}. As expected, for $w\ll 1$ the $AdS_2$ geometry dominates and an IR scaling emerges. By fitting the numerical data, the conductivity can be written as
\be
\sigma(w)={1\over 3}w^2+{1\over 2\sqrt{3}}\left(\delta(w)+{i\over w}\right)+\cdots\sp w\to 0\label{sT0}
\ee
in agreement with the general scaling exponents derived in \cite{scaling}.
We have also included the $\delta$-function in the real part, which is there, due to the $1/w$ pole in the imaginary part.
To finish the numerical analysis of the conductivity, we proceed to plot the absolute value $|\sigma|$ and the argument $\arg\sigma$ as shown in Fig. \ref{fig_T0_b}.
In the left plot, the $1/w$ pole of the imaginary part dominates over the $w^2$ scaling of the real part. In the right plot, the argument takes the value $\arg\sigma=\pi/2$ in the IR, consistent with
\be
\arg\sigma\approx\arctan{\sqrt 3\over 2w^3}\approx{\pi\over 2}-{2w^3\over\sqrt{3}}+\cdots.
\ee
On the contrary in the UV region ($w\gg 1$) the behaviour is determined by the asymptotic $AdS_4$ region
\be \sigma(w)=1,~Arg(\sigma)=0.\ee


\vskip 1cm
\subsection{Temperature versus critical IR scaling of the AC conductivity}
\vskip 1cm

Having understood the zero temperature conductivity, we now introduce a non-vanishing $\tau$, while keeping $\kappa=0$. In this case $\tau$ is the only dimensionless parameter. Therefore, the conductivity depends parametrically only on the dimensionless temperature $\tau$.

\begin{figure}[t]
	\begin{center}
		\includegraphics[scale=0.80]{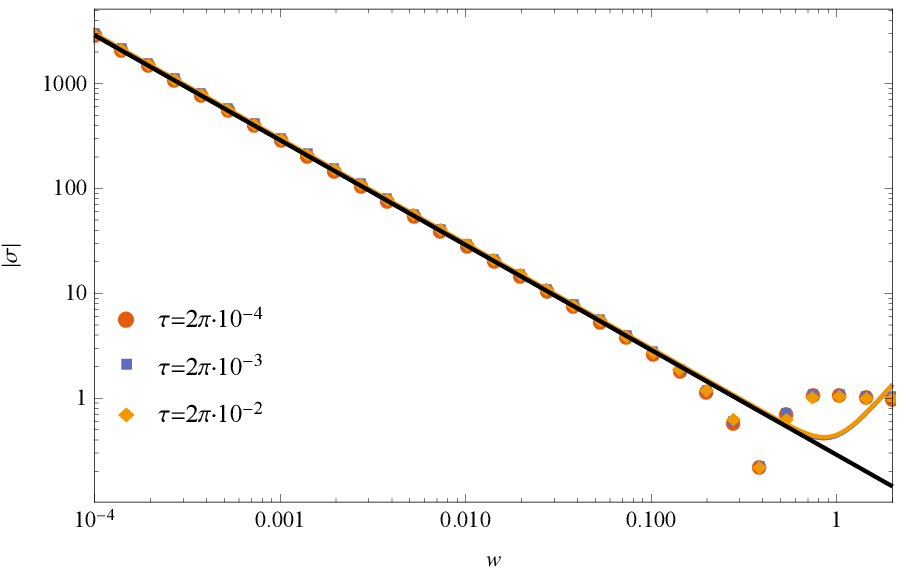}
		\includegraphics[scale=0.80]{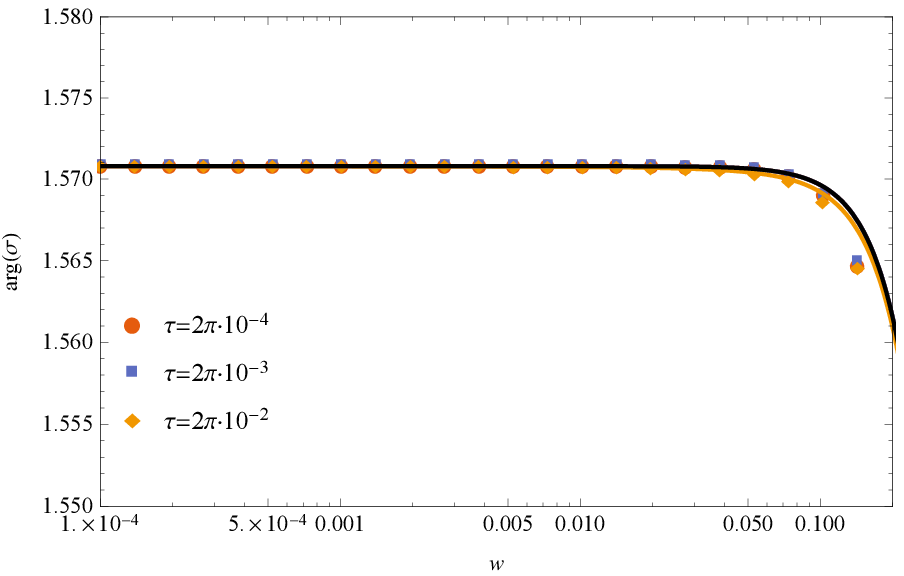}
	\end{center}
	\caption{The absolute value and argument of the conductivity at $\kappa=0$ and at various value of $\tau$. The dots show the numerical data, while the continuous lines are given by (\ref{11}). }
	\label{fig_kzerotsmall2}
\end{figure}

 The equation of motion for the fluctuating gauge field reads (see Appendix \ref{app_fluct})
\be\label{eq_kappa0}
fa_x'' + f'a_x' + F^2\left(\frac{w ^2}{ f} -  z^2\right)a_x=0\,,
\ee
with the blackening factor
\be
f(\rho)=1-\rho^3 + {1\over 4}F^2\rho^3(\rho-1)\,.
\ee
\begin{figure}[t]
	\begin{center}
		\includegraphics[scale=0.80]{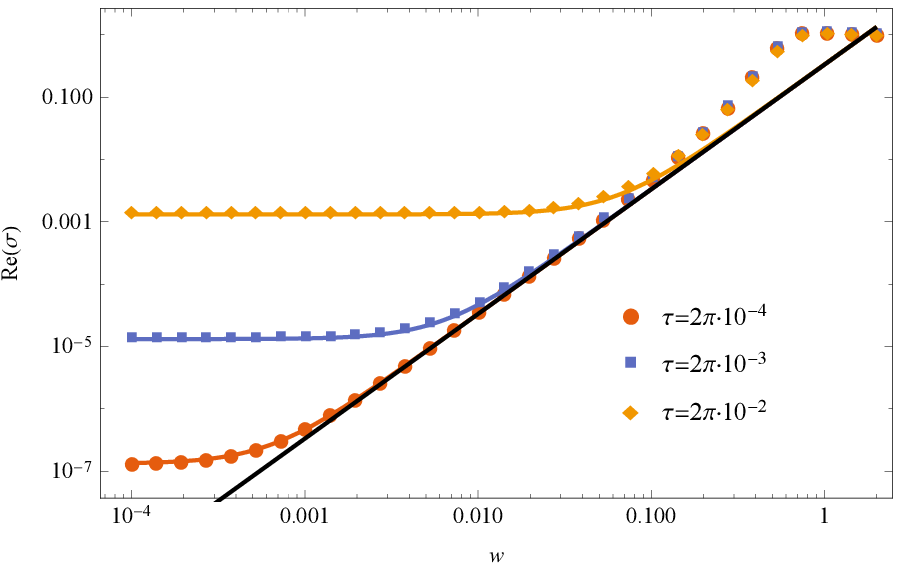}
		\includegraphics[scale=0.80]{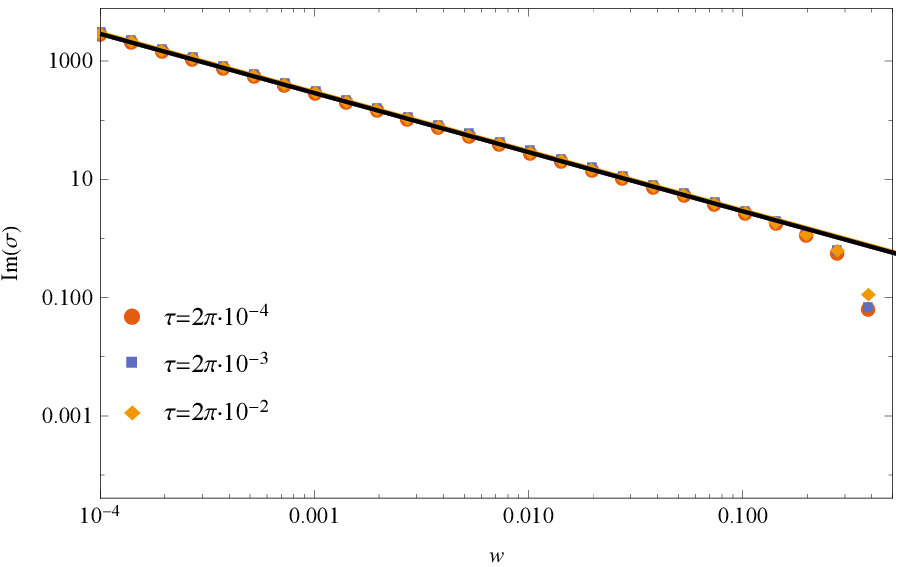}
	\end{center}
	\caption{Finite temperature conductivity at zero momentum relaxation parameter ($\kappa=0$). In the left plot we show the real part for several values of the temperature $\tau$. The $\delta(w)$ in the real part is not drawn. In the right plot the imaginary part is shown. The dots represent numerical data, while the continuous lines are given by \eqref{11}. The black dashed line in the real part corresponds to $Re(\sigma)={1\over 3}w^2$. }
	\label{fig_kzerotsmall}
\end{figure}
The IR conductivity can be studied analytically for $w\ll 1$. For concreteness we show here the result and refer the reader to Appendix  \ref{app1} for the details of the computation. In the regime $w\ll 1$, the conductivity reads
\be
\sigma(w)\approx\sigma_Q+D\left(\delta(w)+{i\over w}\right)\label{10},
\ee
where
\begin{align}
\sigma_Q =\left({12-F^2\over 3(4+F^2)}\right)^2\,,\qquad D ={4F\over 3(4+F^2)}.
 \end{align}
In particular, in the regime of interest ($\tau\ll 1$), the low-frequency conductivity takes the simple form
\be
\sigma(w)\approx {1\over 3}\tau^2 +{i\over 2\sqrt 3 w}.
\ee
Having understood the small frequency analysis, we proceed to solve numerically Eq. \eqref{eq_kappa0}, and show the results  in Fig. \ref{fig_kzerotsmall}. In the left plot, we observe how temperature introduces the constant offset ($\sigma_Q$) to the real part. We shall refer to the regime where this constant value dominates as the temperature-dominated regime. If the condition $\tau\ll w\ll 1$ is satisfied, we notice the emergence of the $AdS_2$ scaling $\sim w^2$. For high enough temperatures, the temperature-dominated regime ``covers" the scaling regime and, thus, the latter is not visible. Given this behaviour, we propose the following form for the low-frequency conductivity
\be
\sigma(w)\approx\sigma_Q+D\left(\delta(w)+i{1\over w}\right)+{1\over 3}w^2\label{11},
\ee
which is shown as continuous lines in figures \ref{fig_kzerotsmall} and \ref{fig_kzerotsmall2}.

In addition, we plot the absolute value and argument of the conductivity for different temperatures in Fig. \ref{fig_kzerotsmall2}. In this case, as it also happens at zero temperature, the $1/ w$ imaginary part of the conductivity always dominates in the IR part of the absolute value of the conductivity. This is easy to see from Eq. (\ref{11}). For the $w^2$ term to ``win" over $D/w$ we need $D\ll w^3$. However, for $\tau\ll 1$, we have $D\approx 0.29\gg w^3$.  Finally, the argument has a similar behaviour to the zero-temperature case at small frequencies.  For $\tau\ll 1$ it reads
\be \arg(\sigma)={\pi\over 2}-{2\over \sqrt 3}{w(w^2+\tau^2)}+\cdots.\ee

\vskip 1cm
\subsection{Momentum dissipation versus IR scaling of the AC conductivity}
\vskip 1cm

The last case to be considered is the general case of finite temperature and momentum relaxing parameter. The system is controlled by the two dimensionless parameters $\tau,\kappa$. For the present case, the gauge field couples to the scalar and metric sector (see Appendix \ref{app_fluct}), therefore we need to solve the system of equations
\begin{eqnarray}
\frac{w F^2 }{\rho^2  f}\left(w \chi -i \kappa  h_t^x\right)+ \left(\rho^{-2}f\chi '\right)' &=& 0\,, \\
-\frac{i \rho^2 F^2w  }{ f }a_x + \frac{i w }{f} h_t^{x'} - \kappa \chi '  &=& 0\,,\\
- h_t^{x'} +\frac{w ^2 F^2 a_x}{ f}  + \left(fa_x'\right)' &=& 0 \,,
\end{eqnarray}
where the blackening factor takes the form
\be
f(\rho)=(1-\rho)(1+\rho+\rho^2-{1\over 2}\kappa^2F^2\rho^2- {1\over 4}F^2\rho^3) \,.
\ee
\begin{figure}[t]
	\begin{center}
		\includegraphics[scale=0.80]{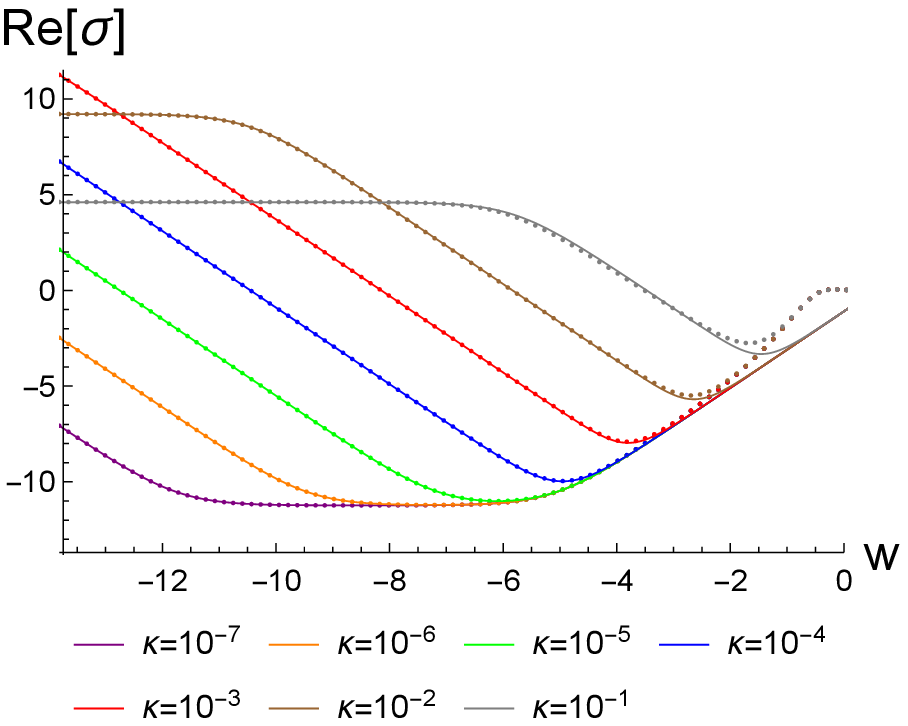}
		\includegraphics[scale=0.80]{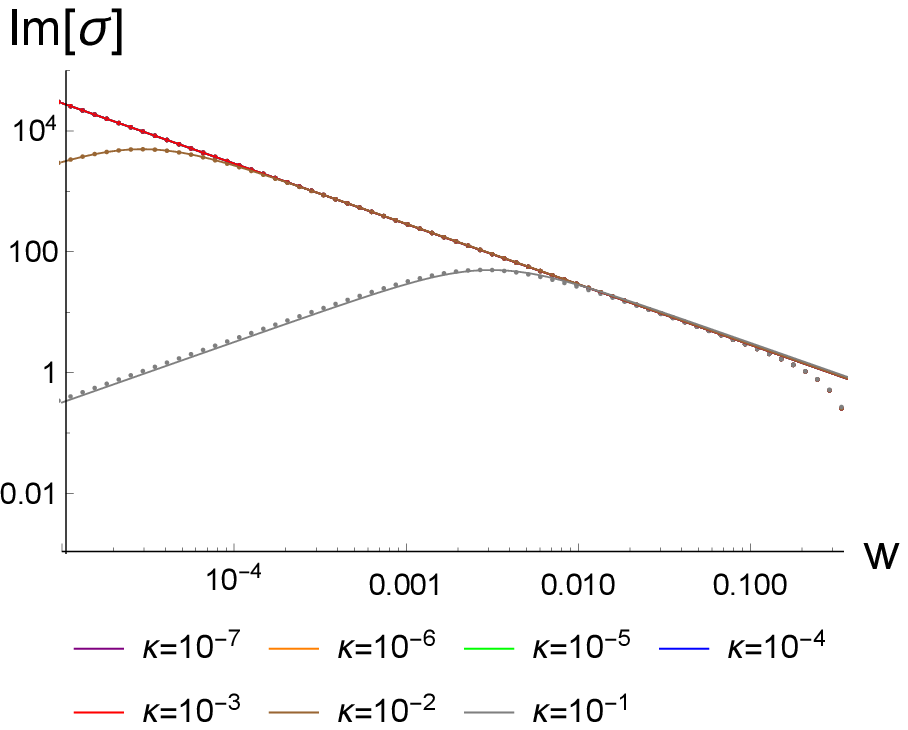}
	\end{center}
	\caption{The real part, imaginary part, absolute value and argument of the conductivity at $\tau=2\pi\cdot10^{-3}$ and for various values of $\kappa$. The dots show the numerical data, while the continuous lines are given by (\ref{expwk}).}
	\label{tcons0}
\end{figure}
In appendix \ref{app2} we solve the fluctuation equation perturbatively for $w\ll F^{-1}$ and $\kappa\ll F^{-1}$. The conductivity in this limit is given by (\ref{drud})
\be
 \sigma(w)\approx{D\over \Gamma-iw }+\sigma_Q,\label{expwk}
 \ee
where
\be
 \sigma_Q =\left({12-F^2\over 3(4+F^2)}\right)^2\sp 	D ={4F\over 3(4+F^2)}\sp
	 \Gamma =\kappa^2 D\;.
	 \ee
Actually, in the  $\kappa\ll 1$ and $\tau\ll 1$ limit, the leading behavior of the coefficients is given by,
\be
D={1\over 2\sqrt 3}\sp \Gamma={\kappa^2\over 2\sqrt 3}\sp \sigma_Q={\tau^2\over 3}.
\ee
Since for $\kappa\ll1,\tau\ll1$ we have $F^{-1}\simeq 0.3$, the approximation is valid in the region of interest ($w\ll 1$).

After the approximate analytic analysis, we solve numerically for the conductivity, and show the results in Fig. \ref{tcons0}. For the computation of the conductivity we fixed $\tau=2\pi\cdot10^{-3}$, and analyse the transport coefficient for several values of $\kappa$. In order to fit the numerical data, in addition to the analytically computed conductivity (Eq. \eqref{expwk}) we add to the real part the power-law $1/3w^2$, and show the function as continuous lines in figure  \ref{tcons0}. In particular, we observe  that for $\kappa=0.1$ the fit is not very good. This is because we are approaching the boundary of the validity region of the formula Eq. \eqref{expwk} ($\kappa\ll F^{-1}\sim 0.3$).
\begin{figure}[t]
	\begin{center}
		\includegraphics[scale=0.80]{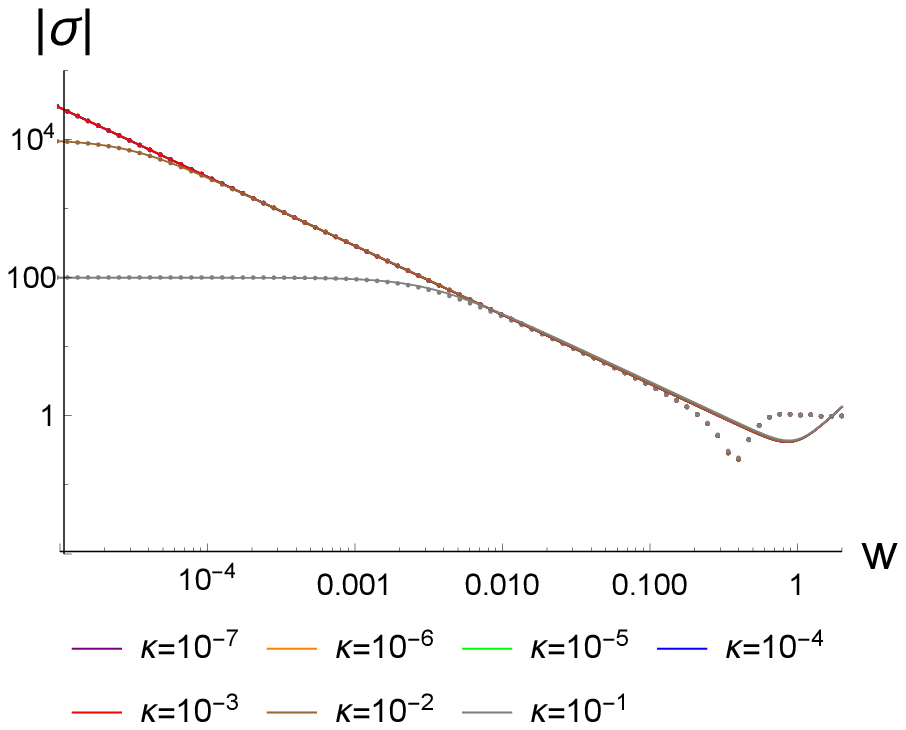}
		\includegraphics[scale=0.80]{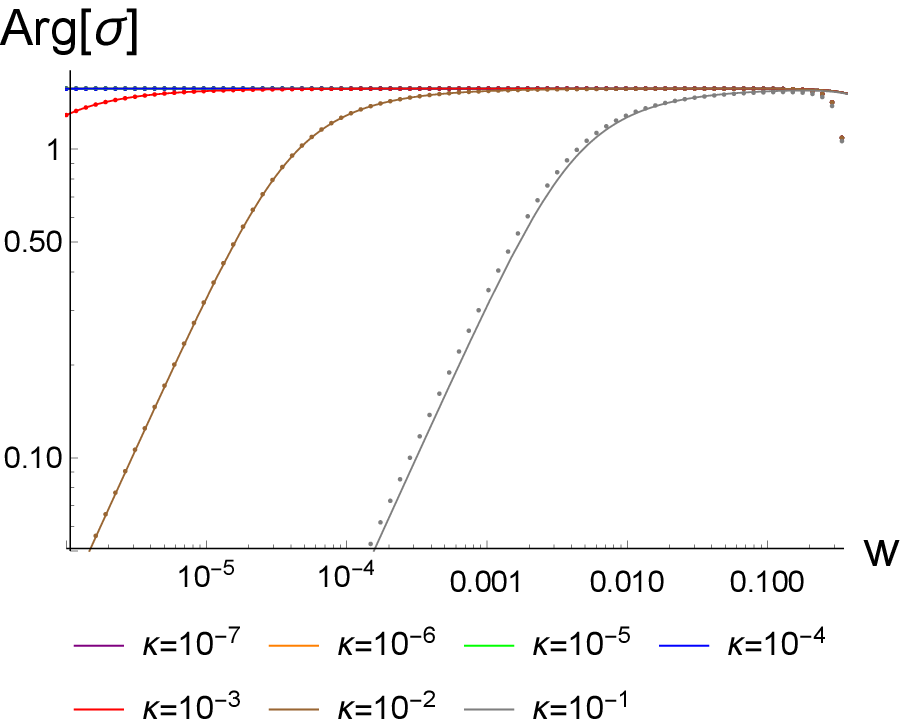}
	\end{center}
	\caption{The real part, imaginary part, absolute value and argument of the conductivity at $\tau=2\pi\cdot10^{-3}$ and for various values of $\kappa$. The dots show the numerical data, while the continuous lines are given by (\ref{expwk}).}
	\label{tcons1}
\end{figure}

We observe that the formula
\begin{equation}
 \sigma(w)\approx{D\over \Gamma-iw }+\sigma_Q +\frac{1}{3}w^2
\end{equation}
approximates well the numerical data as long as $\kappa\ll F^{-1}\sim 0.3$,

We study now the conditions for the scaling of the AC conductivity to be visible. To do so, we write the real part for $\tau\ll1,\kappa\ll1,w\ll1$  as follows
\be
Re[\sigma]\approx\underbrace{\kappa^2 \over {\kappa^4}+12w^2 }_\text{Drude}+\underbrace{{\tau^2\over 3}}_\text{temperature}+\underbrace{{w^2\over 3}}_\text{scaling}.\label{sa1}
\ee
We observe a 'Drude peak' as long as $\kappa\ne 0$, which dominates for small enough frequency. As $w$ increases, either the scaling or temperature terms starts to dominate.
Therefore we divide the analysis in the following two cases\footnote{We  only study the cases where the scaling survives. If either the temperature or the momentum relaxing parameter are large enough,  the critical scaling power-law is no-longer visible in the AC conductivity.}
\begin{itemize}
	\item {\bf Temperature dominated $\lambda \ll \tau \ll 1$:} In this case, as we turn the frequency on, the temperature term in Eq. \eqref{sa1} is the first one to start dominating over the Drude term at frequencies of order $w\sim\kappa/(2\tau)$. Then, as we keep increasing $w$, the scaling term becomes dominant. The temperature-dominated behaviour appears for frequencies $ \kappa/ (2\tau)\ll w\ll \tau$, while the scaling in the conductivity is visible for $ \tau\ll w \ll 1$. In particular, in the left plot of figure \ref{treg} we tuned the parameters to sit within this regime ($\kappa=10^{-7}\,, \tau=2\pi\times 10^{-3}$) and we notice the three well defined regions, Drude, temperature dominated and scaling respectively, in consistency with this classification.
		\begin{figure}[t]
		\begin{center}
			\includegraphics[scale=0.5]{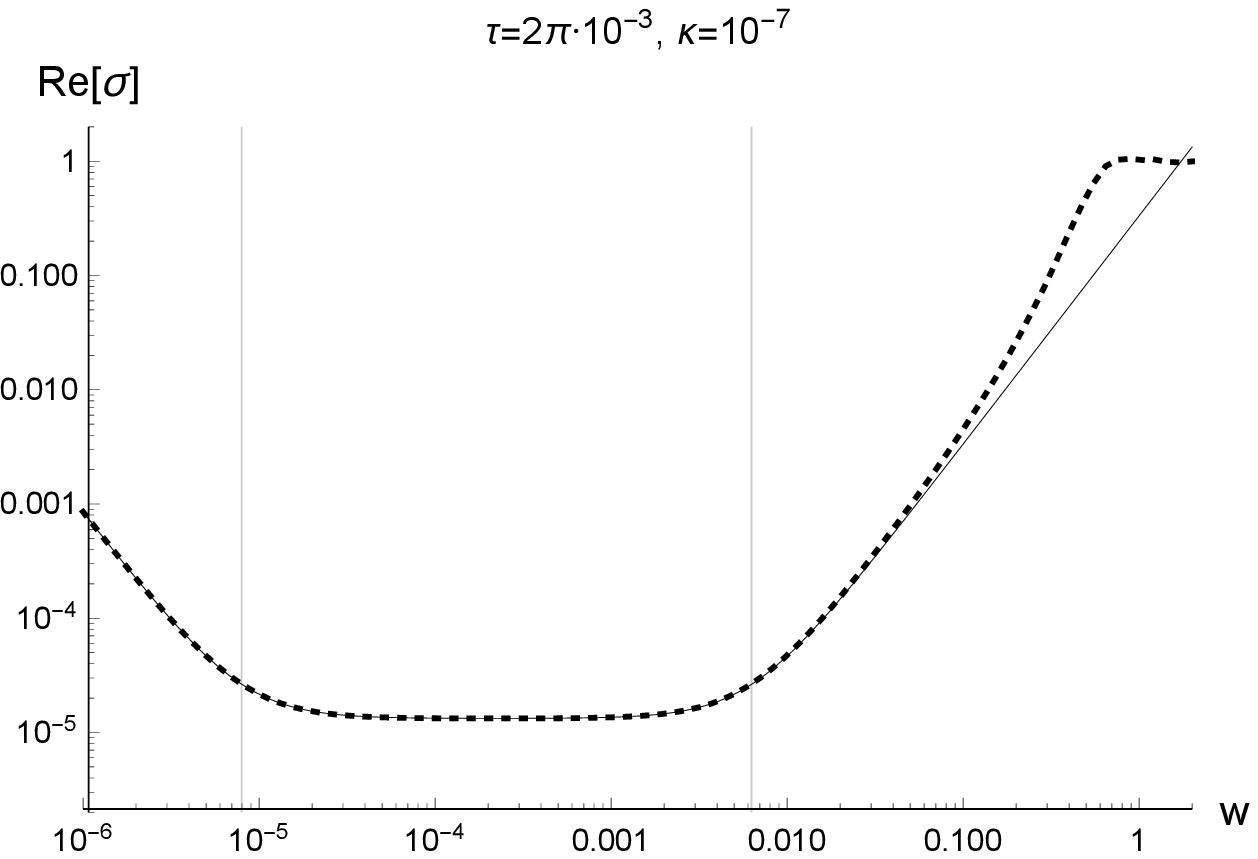}
			\includegraphics[scale=0.5]{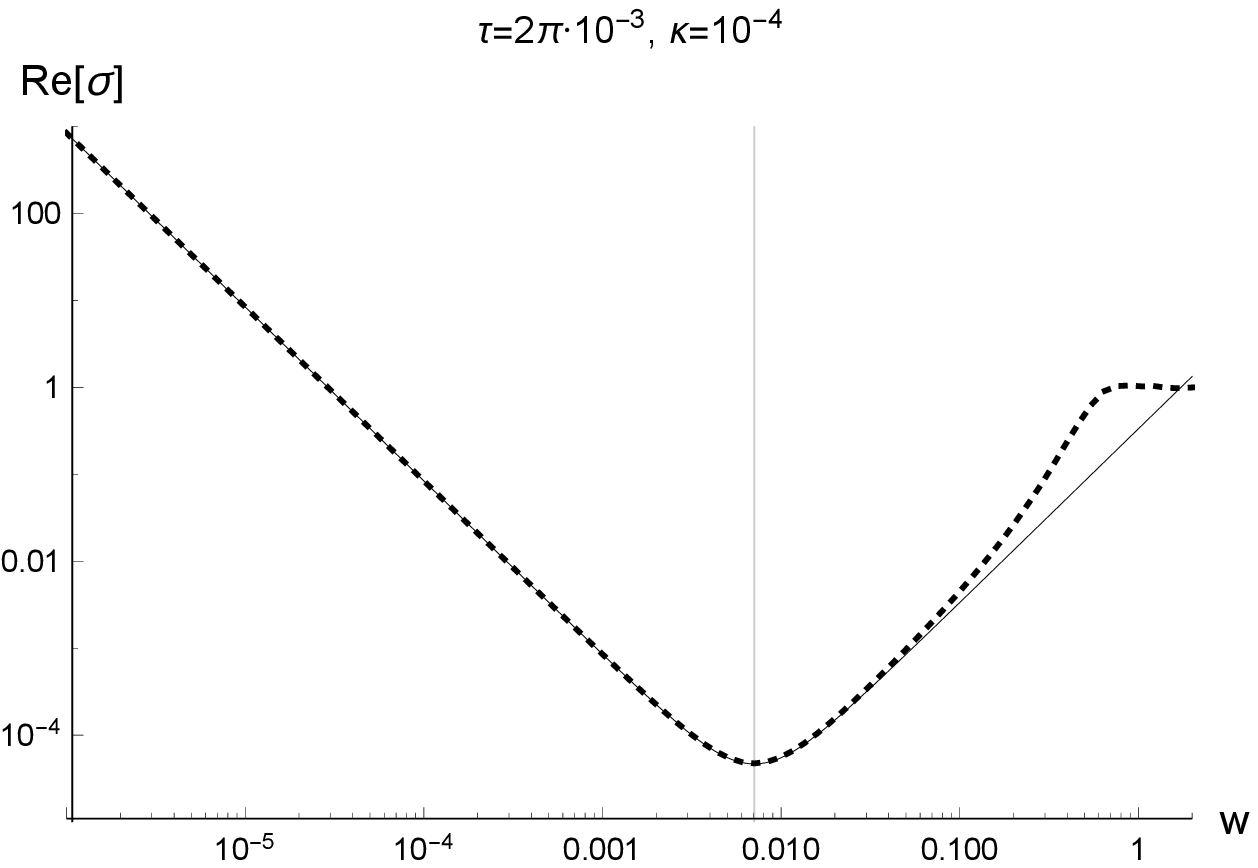}
		\end{center}
		\caption{Real part of the conductivity as a function of the frequency. Left plot shows the conductivity for the case in which $\lambda\ll\tau\ll1$. The Drude peak turns into the ``flat" temperature-dominated behavior, which gives its turn to the scaling behavior $\sim w^2$, before reaching the UV at $w>1$. The vertical lines correspond to ${\kappa\over (2\tau)}$ and $\tau$ from left to right. On the contrary the right plot corresponds to regime $\tau\ll\lambda\ll 1$. The Drude peak shows a transitions directly to the $\sim w^2$ scaling behavior. The dashes show the numerical data, while the continuous line is given by (\ref{sa1}). The vertical line corresponds to $\lambda=\sqrt{\kappa\over 2}$.}
		\label{treg}
	\end{figure}
		\item {\bf Drude dominated $\tau\ll\lambda\ll 1$:} This case is characterized by the 'Drude peak' covering the flat region, but not the scaling regime. In fact, when $w \sim \lambda$ the scaling contribution in the conductivity starts dominating as can be seen in the right plot of Fig. \ref{treg}.	
	
\end{itemize}

The previous  analysis suggests that as long as $ {\tau}\ll 1$ and $\lambda\ll 1$, the critical scaling will be visible within the window
\be
\max\left\{\tau,\lambda\right\}\ll w\ll 1.\label{12}
\ee

Finally, to extract the behaviour of the absolute value and argument of the conductivity we proceed to write the full conductivity as follows
\be
\sigma={D\Gamma\over \Gamma^2+w^2 }+\sigma_Q+{1\over 3}w^2+i{Dw\over \Gamma^2+w^2 }+\cdots\;.
\ee
However, in the region given by Eq. \eqref{12} where the scaling is visible, the conductivity takes the approximate form
\begin{equation}
\sigma\approx {1\over 3}w^2+i{D\over w }+\cdots,
\end{equation}
which automatically implies that the imaginary part will be dominant in the absolute value of the conductivity because $D\approx \frac{1}{2\sqrt{3}}$ and the frequency is $w\ll 0.1$.
On the other hand, the argument at zero frequency vanishes as
\be
 Arg[\sigma]\approx\arctan{2\sqrt 3 w\over \kappa^2},
 \ee
and approaches $Arg[\sigma]\approx \pi/2$ when the frequency is within the values given by the interval \eqref{12}.

\vskip 1cm
\section{Conclusions\label{concl}}
\vskip 1cm

Several strongly coupled holographic theories exhibit an AC conductivity that is a scaling function of the frequency $\omega$ in the IR regime, \cite{cgkkm,scaling}.

We embarked here in a study of how non-zero temperature, and momentum dissipation affect the visibility of scaling in the AC conductivity.
In this paper we have studied perhaps the simplest holographic theory at finite density, a 2+1 dimensional (relativistic) CFT.

Such a theory is known to exhibit an unexpected 1-dimensional scale invariance in the IR at finite density, \cite{liu}. This invariance is intimately tied to the appearance of the AdS$_2$ geometry in the near-horizon region of the near-extremal AdS-Reissner-N\"ordstrom black hole.

We have studied the effects that  temperature and relaxation of momentum has on near-extremal black-holes with $AdS_2$ geometry. We have shown, that as long as the condition
\be
\max\left\{\tau,\lambda \right\}\ll w\ll 1,
\label{conc}\ee
is satisfied, the real part of the electrical conductivity will show the critical scaling behaviour determined by the $AdS_2$ near-horizon geometry.
In (\ref{conc}), $\tau$ is the dimensionless temperature, $\lambda=\sqrt{\kappa/2}$, where $\kappa$ is the dimensionless momentum dissipation coefficient, and $w$ is the rescaled frequency, as defined in (\ref{6}), (\ref{lambda}) and (\ref{w}).
The imaginary part of the conductivity is generically dominated by the  Drude peak in this example.

The next step is to investigate more complex holographic systems that have a a closer resemblance to strange metals, where $\s(\omega)\sim \omega^{-a}$ with $0<a<1$, \cite{scaling}.
We shall investigate the emergence of a scaling regime at finite temperature and momentum dissipation.
In such cases, we also expect that the imaginary part f the conductivity will be dominated by the scaling mechanism like the real part.

\vskip 3cm


\section*{Acknowledgements}\label{ACKNOWL}
\addcontentsline{toc}{section}{Acknowledgements}

We thank B. Gout\'eraux, N.  Hussey, A. Mackenzie, C. Panagopoulos, K. Schalm, D. Van der Marel and J. Zaanen for useful discussions.

 This work was supported in part  by the Advanced ERC grant SM-grav, No 669288, and the Deutsche Forschungsgemeinschaft (DFG, German Research Foundation) under Germany's Excellence Strategy through W\"urzburg-Dresden Cluster of Excellence on Complexity and Topology in Quantum Matter - ct.qmat (EXC 2147, project-id 390858490).
F. P-B acknowledges the Quantum Matter Academy of ct.qmat for support.

\newpage

\appendix
\renewcommand{\theequation}{\thesection.\arabic{equation}}
\addcontentsline{toc}{section}{Appendix\label{app}}
\section*{Appendix}

\section{Equations of Motion}\label{app:EoM}

In this appendix, we list the equations of motion stemming from the action (\ref{eq:bulkaction}).

Taking variations of the action respect to metric, gauge and scalar fields we obtain the following set of equations of motions
\begin{subequations}\label{eom001}
	\begin{align}
	&R_{\mu\nu}= -{3\over L^2}g_{\mu\nu}+{1\over 2}\left(\partial_\mu\phi_1\partial_\nu\phi_1+\partial_\mu\phi_2\partial_\nu\phi_2\right)+{1\over
		2}\left[F^{\;\rho}_\mu\ F_{\nu\rho}-\frac{g_{\mu\nu}}4 F^2\right]\label{eq:1}\\
	&\nabla_\mu F^{\mu\nu}= 0\label{eq:2}\\
	&\nabla_\mu \nabla^\mu \phi_i=0,\;\;\; i=1,2\label{eq:3}
	\end{align}
\end{subequations}
where the covariant derivatives are defined with the Christoffel connection.

We consider static, rotationally-symmetric  (in the $x-y$ plane) solutions, with translational symmetry broken only by the axion fields
\be ds^2=-D(r)dt^2+B(r)dr^2+C(r)dx_idx^i,\;\; \phi_i=kx_i, \;\; A_\m=(A_t(r),0,0,0),\label{ans2}\ee
where $i=x,y$ above.
Substituting the ansatz (\ref{ans2}) into (\ref{eom001}) we obtain the following set of independent equations (the scalar equation of motion are identically satisfied)
\begin{subequations}\label{eom2}
	\begin{align}
	& {6\over L^2}B+L^2{A_t'^2\over 2D}+{B'D'\over 2BD}-{C'D'\over CD}+{D'^2\over 2D^2}-{D''\over D}=0\label{EEQ1}\\
	& -2{C''\over C'}+{C'\over C}+{B'\over B}+{D'\over D}=0\label{EEQ2}\\
	& k^2{B\over C}-{6\over L^2}B+{L^2A_t'^2\over 2D}-{B'C'\over 2BC}+{C'D'\over 2 CD}+{C''\over C}=0\label{EEQ3}\\
	& \left({CA_t'\over \sqrt{BD}}\right)'=0.\label{GFEQ1}
	\end{align}
\end{subequations}
We are interested in asymptotically AdS$_4$ solutions. Under this requirement, the regular solution of (\ref{eom2}) is given by (\ref{2a})-(\ref{3}).

\vskip 1cm
\section{Derivation and analysis of the AC conductivity}\label{app_fluct}
\vskip 1cm

In this appendix,  we derive the equations that determine the AC conductivity in our theory. This is done by deriving the equations of the linear fluctuations around the solutions that perturb the charge density, solving them and then extracting the IR limit of the current-current correlator from the near-boundary expansion.

To study the linear response of the system, we introduce the relevant fluctuating fields, \cite{Kim:2014bza},
\be
\delta A_x=a_x(r)e^{-i\omega t},\;\; \delta g^x_{t}={r^2\over L^2}h^x_t(r)e^{-i\omega t},\;\; \delta\phi_1=\chi(r)e^{-i\omega t}.
\ee
The linearised equations of motion,   for the fluctuations, stemming from the equations (\ref{eom001}),  are
\begin{subequations}\label{fluceq}
	\begin{align}
	&\frac{\omega }{r^2 f(r)}\left(\omega \chi (r)-i k  h_t^x(r)\right)+ \left(r^{-2}f(r)\chi '(r)\right)' = 0 \\
	&\frac{i r^2\omega  A_t'(r)}{ f(r)}a_x(r) + \frac{i \omega }{f(r)} h_t^{x'}(r) - k \chi '(r)  = 0\\
	&A_t'(r) h_t^{x'}(r) +\frac{\omega ^2 a_x(r)}{ f(r)}  + \left(f(r)a_x'(r)\right)' = 0 .
	\end{align}
\end{subequations}

\vskip 1cm
\subsection{Small frequency solution without momentum dissipation\label{app1}}
\vskip 1cm

Starting from (\ref{fluceq}), we set $k=0$ and define the following dimensionless variables
\be \tau=2\pi{T\over \mu}\sp w={\omega\over \mu}\;. \ee
The equations (\ref{fluceq}) can be decoupled to obtain a single equation that governs the fluctuation of the gauge field
\be  fa_x'' + f'a_x' + r_0^2\left(\frac{w ^2}{ f} - \rho^2\right)a_x=0,\ee
where we are using the rescaled radial coordinate
\be \rho~=~{r\over r_0}\ee
and the blackening factor is
\be
f(\rho)=(1-\rho)(1+\rho+\rho^2- {1\over 4}r_0^2\rho^3).\ee

We now change variables by transforming  $a_x$ as follows,
\be a_x=g(\r)Y(\r),~~~ g(\r)=1-{4r_0^2\over 12+3r_0^2}\rho\ee
to obtain
\be (fg^2Y')'+r_0^2{w^2\over f}g^2Y=0.\ee
We now set
\be Y=f(\r)^{iw r_0/f'(1)}X(\r),\ee
to remove the leading behavior at the horizon and obtain an equation for $X(\rho)$
\be X''+\left(\left(1+{iw\over\tau}\right){f'\over f}+{2g'\over g}\right)X'+\left(r_0^2{w^2\over f^2}-{w^2 f'^2\over 4\tau^2 f^2}+{iw f'g'\over  \tau fg}+{iw f''\over 2 \tau f}\right)X=0.\label{eq:fluck0X}\ee

Once the solution for $X(\r)$ is found, then the conductivity is obtained from the near-boundary behavior, and is given by
\be \sigma(w)=-{i\over w r_0}\left(g'(0)+{X'(0)\over X(0)}\right).\ee

We can find a perturbative solution in the IR by expanding $X$ for small $w$ as follows
\be X(\r)=X_0(\r)+wX_1(\r)+w^2X_2(\r)+w^3X_3(\r)+\cdots.\ee
The equations at each order in $w$ are
\bea (fg^2X_0')'&=&0\\
(fg^2X_1')'&=&-i(g^2X_0^2f')'/X_0\\
(fg^2X_2')'&=&-i(g^2X_1^2f')'/X_1-g^2X_0((4\pi \tau)^2r_0^2-f'^2)/f\\
(fg^2X_3')'&=&-i(g^2X_2^2f')'/X_2-g^2X_1((4\pi \tau)^2r_0^2-f'^2)/f\\
&\cdots\\
(fg^2X_n')'&=&-i(g^2X_{n-1}^2f')'/X_{n-1}-g^2X_{n-2}((4\pi \tau)^2r_0^2-f'^2)/f.
\eea

We define the following function,
\be H_n=\int_1^\r(-i(g^2X_{n}^2f')'/X_{n}-g^2X_{n-1}((2 \tau)^2r_0^2-f'^2)/f)~\ud\r.\ee
Then the solution for $X_n$ can be found recursively by
\be X_n=\int_0^\r {H_{n-1}\over g^2 f}~\ud\rho.\ee

The solution for $X_0$ which is regular at the horizon is just a constant $X_0=c$. Then $X_1$ is
\be X_1(\r)=-ic\int_0^\r{2 \tau g(1)^2+g(\r')^2f'(\r')\over f(\r')g(\r')^2}~\ud\r'.\ee

To first order in $w$, the conductivity is given by
\be \sigma(\omega)=-i{g'(0)\over r_0w}+g(1)^2+\mathcal{O}(w)\label{sk0}.\ee
The real part reads
\be Re[\sigma(\omega)]=(2\tau)^2\left({2(\sqrt{(2\tau)^2+3}-2\tau)\over 3(4+2(2\tau)^2-4\tau\sqrt{(2\tau)^2+3})}\right)^2+\mathcal{O}(w^2).\ee

\vskip 1cm
\subsection{The small frequency behavior with weak momentum dissipation}\label{app2}
\vskip 1cm

Starting with the system (\ref{fluceq}) we define the dimensionless variables
\be \tilde{\omega}=\omega r_0\sp \tilde k=kr_0\sp \tilde q=-qr_0^2\label{defs}\ee
and use the radial coordinate
\be \rho={r\over r_0}\ee
to obtain
\begin{subequations}\label{axfluc}
	\begin{align}
	&f\left(fa_x'\right)'+\tilde{q}fh'^x_t+\tilde{\omega}^2a_x=0\label{axbh}\\
	&\rho^2f{(f \rho^{-2}\chi')'}+{\tilde{\omega}^2}\chi-i\tilde{\omega} {\tilde{k}}h^x_t=0\label{axk}\\
	&	i\tilde{\omega }h'^x_t-\tilde{k}f\chi'+i\tilde{q}\tilde{\omega} \rho^2 a_x=0,\label{axcons}
	\end{align}
\end{subequations}
where
\be
f(\rho)=(1-\rho)(1+\rho+\rho^2-{1\over 2}\kappa^2r_0^2\rho^2- {1\over 4}r_0^2\rho^3).\ee

In order to decouple (\ref{axk}), we
define the following functions, \cite{Ge:2010yc},
\be \phi_\pm={h'^x_t\over \r^2}+\tilde qa_x+{C_\pm\over \r}a_x,\label{phipm}\ee
where
\be C_\pm={6\tilde k^2-3\tilde q^2-12\over 8\tilde q}\pm {\sqrt{64\tilde k^2\tilde q^2+(12-6\tilde k^2+3\tilde q^2)^2}\over 8\tilde q}.\label{cpm}\ee
We obtain a decoupled system for $\phi_\pm$
\be (\r^2f\phi_\pm')'+\left({\r^2\omega^2\over f}+\lambda_\pm \r\right)\phi_\pm=0,\label{770}\ee
where
\be \lambda_+={C_+f'+\r(C_-+\tilde q\r)(\tilde k^2-C_+\tilde q\r)\over C_+-C_-}\label{771}\ee
\be \lambda_-={-C_-f'-\r(C_++\tilde q\r)(\tilde k^2-C_-\tilde q\r)\over C_+-C_-}.\label{772}\ee
To first non-trivial order in $\tilde k$ we have
\be \lambda_+=\tilde k^2\left({\rho(-12-3\tilde q^2+4\rho\tilde q^2)\over 12+3\tilde q^2}\right)+\mathcal{O}(\tilde k^4)\label{773}\ee
\be \lambda_-=-{3\over 4}\rho^2(4+\tilde q^2)+\tilde k^2\left({\rho(-24-6\tilde q^2+(36+\tilde q^2)\rho)\over 24+6\tilde q^2}\right)+\mathcal{O}(\tilde k^4)\label{774}.\ee
We also write $f$ as follows
\be f(\r)=f_0(\r)+\tilde k^2 f_1(\r),\label{775}\ee
where
\be f_0(\r)=(1-\r)\left(1+\r+\r^2-{1\over 4}\tilde q^2\r^3\right)\sp f_1(\r)=-{1\over 2}\r^2(1-\r).\label{776}\ee

\begin{itemize}
	\item We start from the equation for $\phi_+$ (\ref{phipm}). Using
	\be \phi_+=\psi f^{i\omega\over f'(1)}\label{777}\ee
	removes the leading behavior at the horizon. Now $\psi$ must be regular at the horizon.
	
	We expand $\psi$ for small $\tilde\omega,\tilde k$ as follows
	\be \psi=\psi_0+\tilde\omega \psi_1+\tilde k^2\psi_2+\mathcal{O}(\tilde\omega^2,\tilde k^4,\tilde\omega\tilde k^2).\label{778}\ee
	The equation for $\psi_0$ is
	\be r^2f_0\psi_0'=c_0\label{779}\ee
	for which regularity at the horizon implies $c_0=0$, hence $\psi_0$ is constant.
	Using this fact we obtain the following equations for $\psi_1,\psi_2$:
	\begin{subequations}\label{780}
		\begin{align}
		&(\r^2f_0\psi_1')'+{i\psi_0\over f_0'(1)}(\r^2f_0')'=0\label{p1}\\
		&(\r^2f_0\psi_2')'+\rho \psi_0 B_1=0\sp B_1={\rho(-12-3\tilde q^2+4\rho\tilde q^2)\over 12+3\tilde q^2}\label{p2}.
		\end{align}
	\end{subequations}
	From (\ref{p1}) we find
	\be \psi_1=i\psi_0\int_1^\rho {1\over f_0}\left({1\over\r^2}-{f_0'\over f_0'(1)}\right)~\ud\rho\equiv i\psi_0 P_1(\rho).\label{781}\ee
	From (\ref{p2}) we find
	\be \psi_2= -\psi_0 P_2(\rho)\sp P_2(\rho)={4(\r-1)\over (12+3\tilde q^2)\r}.\label{782}\ee
	
	\item Now we find a perturbative solution for $\phi_-$. We first use the transformation
	\be \phi_-=gY\label{783}\ee
	with
	\be g={1\over\r}-{4\tilde q^2\over 3(4+\tilde q^2)}\label{784}\ee
	so that the coefficient of $Y$ vanishes at the limit $\tilde\omega\to 0\sp\tilde k\to 0$.
	We use
	\be Y=X f^{i\omega\over f'(1)}\label{785}\ee
	to remove the leading behavior at the horizon. We require that $X$ is regular at the horizon and expand it as follows
	\be X=X_0+\tilde\omega X_1+\tilde k^2X_2+\mathcal{O}(\tilde\omega^2,\tilde k^4,\tilde\omega\tilde k^2).\label{786}\ee
	For $X_0$ we find
	\be \r^2f_0g^2X_0'=c\label{787}\ee
	which, by regularity at the horizon, implies that $X_0$ is constant. Using this fact we obtain the following equations
	\begin{subequations}\label{788}
		\begin{align}
		&(\r^2f_0g^2X_1')'+{iX_0\over f_0'(1)}(\r^2f_0'g^2)'=0\label{x1}\\
		&(\r^2f_0g^2X_2')'+\rho X_0 B_2=0\sp B_2={2\rho\tilde q^2(36+\tilde q^2)(\tilde q^2(4\r-3)-12)\over 27(4+\tilde q^2)^3}\label{x2}.
		\end{align}
	\end{subequations}
	From (\ref{x1}) we obtain
	\be X_1=iX_0\int_0^\rho \left({g(1)^2\over \r^2f_0g^2}-{f_0'\over f_0f_0'(1)}\right)\equiv iX_0 Q_1(\rho).\label{789}\ee
	From (\ref{x2}) we find
	\be X_2= -X_0 Q_2(\rho)\sp Q_2(\rho)={8\tilde q^2(36+\tilde q^2)\over 3(4+\tilde q^2)(-12+\tilde q^2(-3+4\r))^2}.\label{790}\ee
	
	\item We now need to fix the integration constants $X_0,\psi_0$ in terms of the boundary values
	\be a_x^{(0)}=a_x(0)\sp \chi^{(0)}=\chi(0)\sp h_t^{x(0)}=h_t^{x}(0).\label{791}\ee
	The system (\ref{axfluc}) implies the equation
	\be f\r^2\left(\rho^{-2}h'^{x}_t+\tilde q a_x\right)'+{\tilde k^2 h_t^{x}}+{i\tilde k\tilde \omega \chi}=0.\label{zpp}\ee
	Using (\ref{phipm}) and (\ref{zpp}) we find
	\be f\left(\r^2\phi_\pm'-C_\pm a_x'\r+C_\pm a_x\right)=\tilde k^2h^{x}_t+i\tilde k\tilde\omega \chi.\label{792}\ee
	Near the boundary we obtain
	\be \lim\limits_{\r\to 0}(\r^2\phi_\pm')=-C_\pm a_x^{(0)}+\tilde k^2 h_t^{x(0)}+i\tilde k\tilde\omega \chi^{(0)}.\label{nbeq}\ee
	The expansion of $\phi_\pm$ near the boundary is\footnote{there are no logarithms in the expansion; one can check from the solution that $\phi_\pm'$ do not contain any $1/\rho$ terms.}
	\be \phi_\pm=-{W_\pm\over \r}+D_\pm+\cdots\label{793}\ee
	where
	\begin{subequations}\label{WD}
		\begin{align}
		&W_+=\psi_0(i\tilde\omega- {4\tilde k^2\over 12+3\tilde q^2})+\mathcal{O}(\tilde\omega^2,\tilde\omega\tilde k^2,\tilde k^4)\label{794}\\
		&W_-=-X_0+\mathcal{O}(\tilde\omega^2,\tilde\omega\tilde k^2,\tilde k^4)\label{795}\\
		& D_+=\psi_0+\mathcal{O}(\tilde\omega,\tilde k^2)\label{796}\\
		&D_-=X_0\left(-{4\tilde q^2\over 12+3\tilde q^2}+i\tilde\omega {(\tilde q^2-12)^2\over (12+3\tilde q^2)^2}-\tilde k^2 {8\tilde q^2(36+\tilde q^2)\over (12+3\tilde q^2)^3}\right)+\mathcal{O}(\tilde\omega^2,\tilde\omega\tilde k^2,\tilde k^4).\label{797}
		\end{align}
	\end{subequations}
	Then (\ref{nbeq}) implies
	\be W_\pm=-C_\pm a_x^{(0)}+\tilde k^2 h_t^{x(0)}+i\tilde k\tilde \omega \chi^{(0)}+\mathcal{O}(\tilde\omega^2,\tilde\omega\tilde k^2,\tilde k^4),\label{798}\ee
	therefore
	\be \psi_0={-C_+ a_x^{(0)}+\tilde k^2 h_t^{x(0)}+i\tilde k\tilde \omega \chi^{(0)}\over i\tilde\omega-\tilde k^2 {4\over 12+3\tilde q^2}}+\mathcal{O}(\tilde\omega^2,\tilde\omega\tilde k^2,\tilde k^4)\label{799}\ee
	\be X_0={C_- a_x^{(0)}-\tilde k^2 h_t^{x(0)}-i\tilde k\tilde \omega \chi^{(0)} }+\mathcal{O}(\tilde\omega^2,\tilde\omega\tilde k^2,\tilde k^4).\label{7100}\ee
	From (\ref{phipm}) we can solve for $a_x$:
	\be a_x=\r{\phi_--\phi_+\over C_--C_+}\label{7101}\ee
	which implies
	\be a_x'(0)={D_--D_+\over C_--C_+},\label{7102}\ee
	where $D_\pm$ are given in (\ref{WD}). Using also (\ref{cpm}), we obtain the terms relevant to the electric conductivity
	\be \delta a_x'(0)/\delta a_x^{(0)}=i\tilde\omega {(12-\tilde q^2)^2\over (12+3\tilde q^2)^2}-{4i\tilde\omega\tilde q^2\over (12+3\tilde q^2)i\tilde\omega -4\tilde k^2}+\mathcal{O}(\tilde\omega^2,\tilde\omega\tilde k^2,\tilde k^4)\label{7103}\ee
	hence
	\be \sigma(\tilde\omega)= {{\tilde q^2\over\tilde k^2}\over 1 -{(12+3\tilde q^2)\over 4\tilde k^2}i\tilde\omega}+{(12-\tilde q^2)^2\over (12+3\tilde q^2)^2}\label{skt}.\ee
	
	Now using (\ref{defs}) we can write (\ref{skt}) as follows
	\be \sigma(w)={D\over \Gamma-iw}+\sigma_Q,\label{drud}\ee
	where
	\be D={4r_0\over 12+3r_0^2}\sp \Gamma=\kappa^2D\sp \sigma_Q=\left({12-r_0^2\over 12+3r_0^2}\right)^2.\ee
\end{itemize}

\end{document}